%% file: markovpleth.tex
\documentclass{article}[12pt]

\usepackage{amssymb, amsthm, amsmath,graphicx}
\usepackage{a4wide}

\newcommand{\beqn}{\begin{eqnarray}\begin{aligned}}
\newcommand{\eqn}{\end{aligned}\end{eqnarray}}

\usepackage{setspace}

\usepackage{graphicx,pstricks}
\usepackage{pst-node,pst-3d,pst-text,pst-tree}

\begin{document}

\newcommand\zerosplit{\Tdot  \Tdot }
\newcommand\nextstep{{\LARGE $\rightarrow$}}
\psset{linewidth=1pt,yunit=5mm,levelsep=15mm,tnsep=2pt}

\begin{titlepage}

\begin{center}

{\Large {\bf Markov invariants, plethysms, and phylogenetics${}^\ast$}}

\vspace{2em}

J G Sumner$^{1,2}$,  M A Charleston$^{1,4,5}$, L S Jermiin$^{3,4,5}$, and P D Jarvis$^{2,\dagger}$  
\par \vskip 1em \noindent
{\it $^1$School of Information Technologies, $^3$School of Biological Sciences, $^4$Centre for Mathematical Biology, $^5$Sydney Bioinformatics, University of Sydney, NSW 2006, Australia}\\
{\it $^2$School of Mathematics and Physics, University of Tasmania, TAS 7001, Australia}\\
\end{center}
\par \vskip .3in \noindent

\vspace{1cm} \noindent\textbf{Abstract} \normalfont 
\\\noindent
We explore model-based techniques of phylogenetic tree inference exercising Markov invariants. 
Markov invariants are group invariant polynomials and are distinct from what is known in the literature as phylogenetic invariants, although we establish a commonality in some special cases.
We show that the simplest Markov invariant forms the foundation of the Log-Det distance measure.
We take as our primary tool group representation theory, and show that it provides a general framework for analyzing Markov processes on trees.
From this algebraic perspective, the inherent symmetries of these processes become apparent, and focusing on plethysms, we are able to define Markov invariants and give existence proofs.
We give an explicit technique for constructing the invariants, valid for any number of character states and taxa.
For phylogenetic trees with three and four leaves, we demonstrate that the corresponding Markov invariants can be fruitfully exploited in applied phylogenetic studies.

\vfill
\hrule \mbox{} \\
{\footnotesize 
{${}^\ast$This is the ``long version'' that includes an extended introduction, a subsection on mixed-weight invariants, a third appendix on the K3ST model, and a more relaxed pace with additional discussion throughout. The ``short version'' appears in \emph{Journal of Theoretical Biology}, 253:601-615, 2008.}\\
{$^{\dagger}$ Alexander von Humboldt Fellow}\\
{\textit{keywords:} invariants, plethysm, phylogenetics, Schur functions, branching rules}\\
{\textit{email:} jsumner@it.usyd.edu.au}\\
UTAS-PHYS-2007-31
}
\end{titlepage}

\tableofcontents

\input{intro}
\input{sec2}

\input{sec3}
\input{sec4}
\input{conc}
\input{append}

\bibliographystyle{plain}
\bibliography{C:/reference/masterAB}

\end{document}

%% file: intro.tex
\section{Introduction}
\subsection{Background}
Molecular phylogenetic methods aim to infer the past evolutionary relationships of organisms from present day molecular data such as nucleotide sequences. 
Progress is made by making astute assumptions about the evolutionary process, which simplify the problem into a mathematical form, while retaining much of the structure motivating the biological question at hand. 
This process of mathematical modeling is essential if informed inferences from observed data sets are to be made.

The most significant simplification made in phylogenetic models is that the evolutionary change of the molecular units is assumed to progress by mutation under environmental influences and the Darwinian effects of selection are ignored. 
Another overriding simplification, featuring in all the popular models, is that the effect of mutations is modelled as a stochastic (random) process assumed to be Markov. Also, it is often assumed that any given site in a molecular sequence evolved independently of the other sites, and the probability of mutation at each site is identically distributed (known together as the IID assumption). Although the IID assumption is known \textit{not} to hold in many cases \cite{lockhart1998}, we will assume throughout that IID holds, and defer modification of the results presented here to this more general case.

Much progress in phylogenetic inference has been achieved in recent years with the use of sophisticated mathematics, probability and statistical theory, and the advent of powerful computing techniques. A general rule that rates the scientific credence of a phylogenetic method is that \textit{model based} techniques are preferred. In particular, some recent work has focused on the elucidation of the implicit model assumptions of popular methods such as Neighbor-Joining \cite{bryant2005,steel2006} and Maximum Parsimony \cite{tuffley1997}. This type of analysis is an essential part of the scientific justification because otherwise it is not exactly clear what is being estimated in the statistical sense. Without such a framework the biologist is left without any information regarding the confidence in the inference produced.

An overlying difficulty in phylogenetic tree inference is that the number of possible trees is vast, and the space of trees is non-Euclidean; hence it is not clear how one should proceed in searching through it. It is normal to begin with a candidate tree and then consider each of its neighbouring trees (under a given adjacency rule) and choose the new tree as that with the best score. 
There is a range of available tree perturbation types such as ``prune and regraft'' or Nearest Neighbour Interchange, which define these adjacencies. Which type is preferable is a matter of ongoing debate \cite{charleston2001,hordijk2005} and such heuristic techniques sometimes find only locally optimal solutions. 
In this paper we will not discuss the problems associated with large trees, but consider how small trees may be built under general model assumptions. 
We give a general framework for constructing small trees, which can then be used as a springboard for building larger trees using techniques such as `quartet puzzling' \cite{strimmer1996} or supertree methods (for arbitrarily sized subtrees) \cite{bininda2004,wilkinson2006}.

Due to its importance for calculating divergence times of lineages, the rate of mutation present in models of evolution is of central importance in phylogenetics. There are several well-known limitations of the standard models involving the rate of mutation on a phylogenetic tree. For instance, the IID assumption is almost always violated by the existence of site-to-site rate variation \cite{pagel2004} and by the existence of invariable sites \cite{lockhart1996}. Other issues include non-stationary processes of evolution (which leads to `compositional heterogeneity' \cite{jermiin2004}), `pattern heterogeneity' (where the pattern of substitutions varies across the sites \cite{pagel2004}) and `heterotachy' (differential rates across the tree) \cite{lockhart2006}. Ignoring the invalidity of the simple models when such assumptions are violated leads to model mis-specification \cite{steel2005} and (potentially) incorrect tree inference.

An issue for any inference technique is that of `consistency'; where consistency is always with respect to an explicit or implicit model (or family of models) of sequence evolution. Statistical consistency requires that if the data set is sampled from a distribution generated under the model assumptions, then the inference method tends to the correct answer 100\% of the time as the size of the data set (length of the sequences) tends to infinity. For example Felsenstein \cite{felsenstein1978} showed that Maximum Parsimony (MP) is statistically \textit{inconsistent} (with all but a small family of models \cite{steel2002}). 

As exemplified by the first three chapters of the recent review book \cite{gascuel2005}, the statistically consistent, model based phylogenetic methods can be placed into three categories: Minimum Evolution (ME) and distance based methods, Maximum Likelihood (ML), and Bayesian methods. ME proceeds by defining a (model based) matrix of pairwise distances between the molecular sequences, and then minimizes the total tree length across the space of possible trees subject to some statistical criteria such as least squares (see Chapter~1 of \cite{gascuel2005}). ML proceeds by maximizing the `likelihood' of the observed data set across the set of possible trees and models of evolution \cite{felsenstein2004,gascuel2005}. Bayesian methods proceed using Bayes' theorem to calculate a posterior distribution on the space of possible trees given a prior distribution (usually uniform--which is an issue in itself as this does not correspond to any evolutionary model of tree generation \cite{yang2006}). For each of these methods the underlying model assumptions are explicit, and current research efforts revolve around implementing these methods under expanded assumptions and/or in a computationally efficient manner.

Another desirable feature of any phylogenetic method is that the model on which it is based should be defined by as few numerical parameters as possible. The issue of scientific content of a model and parameter counts is discussed by Steel \cite{steel2005} in relation to the effectiveness of MP vs ML, where it was stated that the ``predictive power of the theory\ldots tends to be drowned out in a sea of parameter estimation''. This is a fundamental problem in model selection for biological inference, and corresponds to what is known as the bias/variance trade off of parameter estimation \cite{burnham2002}; which in turn equates to the problem of ``overfitting" or ``underfitting" a data set. From an information theoretic perspective, a given data set contains only so much information from which the numerical parameters of a model may be estimated. A model with many parameters may fit the data very well, in that the parameter estimates may be close to their true values, but the corresponding variances will be large because there are relatively few data points. On the other hand, the variance of the estimates of a model with very few parameters will be smaller as there are many data points to estimate each parameter, but in this case the model runs the risk of being badly mis-specified, so that the parameter estimates may be biased. In this light, the `covarion' model \cite{penny2001} deals with the effects of invariable sites whilst introducing only one extra parameter, and the `gamma' model \cite{yang1994} accounts for site-to-site rate variation, with only an additional two parameters. Other methods for coping with heterotachy, rate variation and pattern heterogeneity include the partitioning of data sets and mixture models \cite{pagel2004}. However, all of these methods suffer because, in the general case, the models must include an individual rate matrix (containing up to twelve parameters) and an edge length parameter for each and every edge of the phylogenetic tree. In \cite{posada2004} it was recently noted that the task of phylogenetic tree inference often lies in a region where there are more parameters than data points.

To reduce the number of parameters in phylogenetic models, the evolutionary process is usually assumed to be stationary and reversible, the rate matrices are assumed to have a certain form (such as the Jukes-Cantor model with one parameter, or the Kimura models with two or three parameters), and each edge of the phylogenetic tree is assigned the same rate matrix (for details on these assumptions, see \cite{bryant2005b,jermiin2008}).  
To accommodate non-stationary processes and associated compositional heterogeneity, it becomes necessary to introduce many more parameters into the model. In this circumstance it then becomes desirable to use a technique based on a general model but without the need to estimate the numerical parameters. In this light, a matrix of Log-Det pairwise distances combined with the Neighbor-Joining algorithm \cite{lockhart1994} achieves statistically consistent tree inference under the assumption of a general model. However this technique has its own shortcomings as distance methods only consider pairwise sequence alignments, ignoring much of the information available in the data set, and has problems with model mis-specification \cite{sumner2006}, and the statistical properties of the Log-Det are not exactly known \cite{gu1996}. A recently presented method \cite{jayaswal2007} fits a very general model, but clearly will have issues with over-parameterization and computational requirements.

In summary, the desirable features of a given phylogenetic method are that it is based on a general model of sequence evolution, it is statistically consistent with a family of known models, and the number of parameters to be estimated is minimal.

\subsection{Markov invariants}

In this work we introduce the use of mathematical representation theory to the problem of phylogenetic inference (further background to the results is presented in the PhD thesis \cite{sumner2006a}). We define `Markov invariants' and show that these functions, when evaluated on sequence data, can be put to work in the problem of phylogenetic tree inference under rather general model assumptions. 

Markov invariants are distinct from what is known in the literature as `phylogenetic invariants' \cite{cavender1987,evans1993,lake1987,steel1993}. 
Markov invariants are a particular case of group invariant functions \cite{olver2003} and are hence more constrained by definition than phylogenetic invariants. 
Some Markov invariants are simultaneously phylogenetic invariants, but the reverse is not true in general. 
The structure of Markov invariants is more akin to that of the Log-Det function \cite{lake1994,lockhart1994}, which is constructed using the simplest example of a Markov invariant, yet it is not a phylogenetic invariant. 

The appeal of this approach is that Markov invariants do not assume any particular rate matrices or edge length parameters on the phylogenetic tree. 
Broad conditions of molecular evolution are thus accommodated, incorporating arbitrary substitution rates, non-stationary and time-inhomogeneous processes, heterotachy, and arbitrary pattern heterogeneity across the tree. 
Further, Markov invariants satisfy certain algebraic relations for particular phylogenetic trees, and can provide a novel method of tree inference.

This approach to phylogenetic tree inference satisfies the desirable features given in the summary above. 
That is, Markov invariants are valid for a general model of sequence evolution, statistical consistency is assured, and only a few parameters need to be estimated.

In particular, for the quartet case, we give a tree inference routine, valid for these inclusive conditions, optimizing over only one parameter.
It is hoped that, with additional understanding, this technique can be extended to larger trees.
This will result in phylogenetic tree inference methods, valid for general models, that make use of only a few parameters.
Such a possibility is very attractive, as all of the data is utilized, and a general model may be assumed with the risk of overfitting significantly reduced.

In this paper we outline the theoretical background required to understand the derivation of Markov invariants. 
This will necessitate, in \S\ref{sec:sec1}, an excursion into elementary measure theory on finite sets, and the construction of `phylogenetic tensors'. 
In \S\ref{sec:Groups} we analyse certain groups affiliated with the Markov process, and review standard results from group representation theory. 
This section concludes with a derivation of existence conditions for Markov invariants.  
In \S\ref{sec:markovinv} we report on the structure of Markov invariants for phylogenetic trees with three and four leaves, and give examples of how they can be incorporated into practical phylogenetic analyses. 

%% file: sec2.tex
\section{Measure theory, the Markov semigroup, and phylogenetic tensors}\label{sec:sec1}

In \S\ref{subsec:measures} we collect some basic properties of measures on finite sets, justifying the use of tensor product spaces in the context of Markov processes on phylogenetic trees. The results are rather elementary, but ultimately necessary to place the subsequent discussion on its proper footing. See, for example, \cite{halmos1974} for an introduction to measure theory. In \S\ref{subsec:randomvariables} we use generating function techniques to calculate expectation values of various random variables (and functions thereof) associated with phylogenetic data sets. We give a simple example and show how to compute its unbiased estimator. We define the `Markov semigroup' for the general time-inhomogeneous process (\S\ref{subsec:markovsemi}), construct `phylogenetic tensors' (\S\ref{sec:phylotens}), and, finally, define Markov invariants (\S\ref{subsec:MarkovInvariantsDef}).

\subsection{Probability measures on finite sets}\label{subsec:measures}

Consider a finite set labelled by natural numbers, $ K\!=\!\{1,2,\ldots,k\}$. A \textit{probability measure} on $K$, is a function $\mu\!:K\rightarrow [0,1]$, such that, for any proper subset $A\!\subset\! K$ and any sequence $A_1,A_2,\ldots$ of pairwise disjoint subsets, the following conditions hold:
\beqn
\mu(\varnothing)&=0,\\
\mu(A)&<1,\\
\mu\left(\bigcup_{i} A_i\right)&=\sum_{i}\mu(A_i),\nonumber\\
\mu(K)&=1.
\eqn 
We denote the set of probability measures on $K$ as $\mathcal{M}(K)$.
It follows from the third condition that for $1\!\leq\! i\!\leq \!k$ the measures, $\delta_i(A)\!=\!1$ if $i\in A$ and 0 otherwise, form a basis such that
\beqn
\mu=\sum_{i=1}^k \mu_i\delta_i,\nonumber
\eqn
for all $\mu\in\mathcal{M}(K)$ with $\mu_i\!:=\!\mu\left(\{i\}\right)$. This definition is equivalent to the usual requirement of a probability distribution on a finite set:
\beqn
\sum_{i=1}^{k}\mu_i=\mu\left(\bigcup_{i=1}^k\{i\}\right)=\mu(K)=1.\nonumber
\eqn

In phylogenetics the data sets under consideration are aligned sequences of molecular units. For example, in the case of DNA made up of the four nucleotides adenine, cytosine, guanine, thymine, we would have $K\!=\!\{A,C,G,T\}$, $k\!=\!4$ and write $K\!=\!\{1,2,3,4\}$. However, the results presented here and in \S\ref{sec:Groups} are valid for \emph{any} $k$. In \S\ref{sec:markovinv} we will concentrate on cases relevant to phylogenetics and investigate the Markov invariants for $k\!=\!2,3$ and 4.

In this work we do not consider the problem of aligning the sequence data, and assume throughout that the `true' alignment (without gaps) can and has been found (where truth is relative to the modelling process). Under this circumstance, it becomes necessary to consider the direct product of $K$ with itself $m$ times:
\beqn
K^m:=\times^mK=K\times K\times\ldots\times K\nonumber
\eqn
with $|K^m|\!=\!k^m$. 
Exactly as above, for any proper subset $E\subset K^m$ and any sequence of pairwise disjoint subsets $E_1,E_2,\ldots $, a probability measure, $\mu\in\mathcal{M}(K^m)$, must equivalently satisfy
\beqn
\mu(\varnothing)&=0,\nonumber\\
\mu(E)&<1 ,\\
\mu\left(\bigcup_{i} E_i\right)&=\sum_{i}\mu(E_i),\\
\mu(K^m)&=1.
\eqn
Given that under a measure unions decompose into summations, it follows that we have the tensor product: 
\beqn
\mathcal{M}(K^m)=\otimes^m\mathcal{M}(K):=\mathcal{M}(K)\otimes \mathcal{M}(K)\otimes\ldots\otimes\mathcal{M}(K).\nonumber
\eqn
Concretely, any subset of $K^m$ can be expressed as a union of disjoint subsets of the form
\beqn
A_1\times A_2\times\ldots\times A_m,\nonumber
\eqn
with $A_1,A_2,\ldots ,A_m\subseteq K$. A basis for $\otimes^m\mathcal{M}(K)$ is then, for $1\leq i_1,i_2,\ldots ,i_m\leq k$, 
\beqn
\delta_{i_1}\!\otimes\delta_{i_2}\!\otimes\!\ldots\!\otimes\delta_{i_m}(A_1\times A_2\times\ldots\times A_m)
:=\delta_{i_1}(A_1)\delta_{i_2}(A_2)\ldots\delta_{i_m}(A_m),\nonumber
\eqn 
with $\delta_{i_1}(A_1)\delta_{i_2}(A_2)\ldots\delta_{i_m}(A_m) = 1$ if $\{i_1\}\!\times\!\{i_2\}\!\times\!\ldots\!\times\!\{i_m\}\in A_1\times A_2\times\ldots\times A_m$ and 0 otherwise.
We index the elements $\{i_1\}\!\times\!\{i_2\}\!\times\!\ldots\!\times\!\{i_m\}$ as 
\beqn
I=i_1i_2\ldots i_m,\nonumber
\eqn
and write
\beqn
\mu_I\equiv \mu_{i_1i_2\ldots i_m}:=\mu(\{i_1\}\!\times\!\{i_2\}\!\times\!\ldots\!\times\!\{i_m\}).\nonumber
\eqn
We refer to $m$ as the \emph{rank} of the tensor $\mu$.

Previously the authors JGS and PDJ have presented probability distributions on phylogenetic trees in a tensor product formalism motivated from analogies to quantum physics \cite{jarvis2005,sumner2005}. The formulation presented above places this construction on its proper measure-theoretic footing\footnote{We are indebted to Michael Baake for drawing our attention to this.}. In \S \ref{sec:phylotens} we will relate a given (Markov) model of evolution on a phylogenetic tree with $m$ leaves, to a unique rank $m$ tensor $P\in\otimes^m\mathcal{M}(K)$.

\subsection{Random variables, generating function, expectation values and estimators}\label{subsec:randomvariables}
Any data set considered in a phylogenetic study is necessarily of finite extent, and we suppose that it is a sample drawn from some unknown distribution. We wish to define expectation values of such data (or events) and functions thereof. Throughout we will assume the IID assumption holds, so that we need only consider the distribution of a single random variable. The probability of observing a particular state at a given site will be identical for all the other sites.

For a set of $m$ aligned sequences of length $N$, define a \textit{pattern} to be the (ordered) set of states read across the $m$ sequences at a particular site in the alignment. That is, a pattern takes the form $I\!=\! i_1i_2\ldots i_m$, where $i_a$ is the character state in the $a^{th}$ sequence. Define the random variable $X$ as the pattern observed at a given site. A probability distribution for $X$ can be specified using a probability measure $\mu\in\otimes^m\mathcal{M}(K)$:
\beqn\label{alignEvol}
\mathbb{P}[X\!=\! i_1i_2\ldots i_m]=\mu_{i_1i_2\ldots i_m}.
\eqn
For a sequence of finite length $N$, define $Z$ as the random variable that counts the number of occurrences of each pattern $I\!=\! i_1i_2\ldots i_m$ in the alignment, so that
\beqn
Z=(Z_I)=(Z_{i_1i_2\ldots i_m})_{1\leq i_1,i_2,\ldots,i_m\leq k},\nonumber
\eqn 
and $\sum_{I\in K^m}Z_I\!=\!N$.
Assuming that each site in the alignment is identically and independently distributed as (\ref{alignEvol}), it follows that $Z$ is multinomially distributed under the measure $\mu$:
\beqn
\mathbb{P}[Z\!=\!z;N]=\prod_{I\in K^m}\frac{N!}{z_I!}\mu_I^{z_{I}}.\nonumber
\eqn
This expresses, under the assumptions of $\mu$, the probability of observing within the alignment of $m$ sequences the specific number of occurrences of each of the possible character patterns $Z=z$.

When we describe Markov invariants, we will need to discuss expectation values of the random variable $Z$ and functions thereof. For any function $\phi$, the expectation value with respect to the measure $\mu$ is defined as 
\beqn
E[\phi(Z)]:=\sum_{z}\phi(z)\mathbb{P}[Z\!=\!z;N],\nonumber
\eqn
with the summation over all $z$ such that $\sum_{I\in K^m}z_I\!=\!N$.

Remembering that $Z$ follows a multinomial distribution, it is in practice necessary to use generating function techniques in order to calculate these expectation values. The generating function on the formal variables $s=(s_I)=(s_{i_1i_2\ldots i_m})_{1\leq i_1,i_2,\ldots,i_m\leq k}$ of the multinomial distribution is 
\beqn\label{generatingfunction}
G(s):=E[e^{(s,Z)}]=\left(\sum_{I\in K^m}\mu_{I}e^{s_I}\right)^N,
\eqn
with
\beqn
(s,Z):=\sum_{I\in K^m}s_{I}Z_{I}.\nonumber
\eqn
From the properties of the exponential function and the commutivity of differentiation and expectation,
\beqn
\left.\frac{\partial G(s)}{\partial s_{i_1i_2\ldots i_m}}\right|_{s=0}=E[Z_{i_1i_2\ldots i_m}].\nonumber
\eqn
Using the above closed form of the generating function, an elementary calculation returns
\beqn
E[Z_{i_1i_2\ldots i_m}]=N\mu_{i_1i_2\ldots i_m},\nonumber
\eqn
as of course would be expected. This can be extended to find the expectation of any function of $Z$:
\beqn
E[\phi(Z)]=\left.\phi\left(\frac{\partial }{\partial s}\right)G(s)\right|_{s=0}.\nonumber
\eqn 

As a concrete example, take $m\!=\!2$ and consider the case $\phi(Z)=Z_{44}^2-Z_{12}Z_{13}$. From the linearity of the expectation values we have
\beqn
E[Z_{44}^2-Z_{12}Z_{13}]=E[Z_{44}^2]-E[Z_{12}Z_{13}],\nonumber
\eqn
so  we can consider each term in turn. Taking derivatives of the closed form of the generating function gives
\beqn
E[Z_{44}^2]=N(N-1)\mu_{44}^2+N\mu_{44}\nonumber
\eqn
and
\beqn
E[Z_{12}Z_{13}]=N(N-1)\mu_{12}\mu_{13}.\nonumber
\eqn
Thus, in this case, the expectation value of $\phi$ is
\beqn
E[\phi(Z)]=N(N-1)(\mu_{44}^2-\mu_{12}\mu_{13})+N\mu_{44}.\nonumber
\eqn

Given a (possibly unobservable) random variable $\theta$, an \textit{estimator} is another random variable which is a function of observable quantities such that its expectation value somehow approximates $\theta$. The \emph{bias} of an estimator $\widehat{\theta}$ is defined as the difference
\beqn
b(\hat{\theta})=E[\widehat{\theta}]-E[\theta],\nonumber
\eqn
allowing for $\theta$ to simply be a constant so that $E[\theta]=\theta$. An \textit{unbiased} estimator is simply an estimator with bias equal to zero. For example, a short calculation reveals that the unbiased estimator of $\phi(\mu)$ above is  
\beqn
\frac{\phi(Z)-Z_{44}}{N(N-1)}.\nonumber
\eqn

In general, if $\phi$ is polynomial, computing an unbiased form is a straightforward matter of solving a sequence of difference equations.
When it comes to discussing estimators for Markov invariants, we will show that unbiased forms can easily be defined. 
However, we will note that explicit computation is difficult due to a required change of basis.

\subsection{The Markov semigroup}\label{subsec:markovsemi}

A stochastic process can be described by introducing a time-dependent random variable $X(t)$. A crucial component of the subsequent discussion will be that the time evolution of the corresponding probability distribution can be viewed as a linear mapping upon a vector space. Presently we will establish the conditions for a \textit{Markov} process, and show that such a process satisfies the desired property. See, for example, \cite{isoifescu1980} for an equivalent derivation.

Consider a time-dependent, finite-state random variable, $X(t)$, taking on values in $K$, any set of times $t_1<t_2<\ldots<t_n<t$, and the joint distribution of $X$ across those times: 
\beqn
\mathbb{P}[X(t_1)\!=\! i_1,X(t_2)\!=\! i_2,\ldots,X(t_n)\!=\! i_n,X(t)\!=\!i].\nonumber
\eqn
The distribution of $X$ at the particular time $t$ is given by the marginal,
\beqn
\mathbb{P}[X(t)\!=\!i]=\sum_{1\leq i_1,i_2,\ldots,i_n\leq k}\mathbb{P}[X(t_1)\!=\!i_1,X(t_2)\!=\!i_2,\ldots,X(t_n)\!=\!i_n,X(t)\!=\!i],\nonumber
\eqn
and this can be re-expressed by invoking the conditional distribution:
\beqn
\mathbb{P}[X(t)\!=\!i]=\sum_{1\leq i_1,i_2,\ldots,i_n\leq k}\mathbb{P}&[X(t)\!=\!i|X(t_1)=i_1,X(t_2)\!=\!i_2,\ldots,X(t_n)\!=\!i_n]\\
&\hspace{15mm}\cdot\mathbb{P}[X(t_1)\!=\!i_1,X(t_2)\!=\!i_2,\ldots,X(t_n)\!=\!i].\nonumber
\eqn
The simplest stochastic process is the process for which the probability of a transition to a new state at a given time is independent of the states at all preceding times (such as tossing of a coin--the \textit{Bernoulli process}). A \textit{Markov process} can be seen as the next simplest case where the probability of a transition is independent of all but the state at the most recent time. Thus, for a Markov process the conditional distribution satisfies
\beqn
\mathbb{P}[X(t)\!=\!i|X(t_1)\!=\!i_1,X(t_2)\!=\!i_2,\ldots,X(t_n)\!=\!i_n]=\mathbb{P}[X(t)\!=\!i|X(t_n)\!=\!i_n].\nonumber
\eqn
This implies that the marginal distribution of $X$ at the time $t$ is
\beqn
\mathbb{P}[X(t)\!=\!i]=\sum_{1\leq i_n\leq k}\mathbb{P}[&X(t)\!=\!i|X(t_n)\!=\!i_n]\\
&\cdot\sum_{1\leq i_1,\ldots, i_{n-1}\leq k}\mathbb{P}[X(t_1)\!=\!i_1,X(t_2)\!=\!i_2,\ldots,X(t_n)\!=\!i_n]\\
=\sum_{1\leq i_n\leq k}\mathbb{P}[&X(t)\!=\!i|X(t_n)\!=\!i_n]\mathbb{P}[X(t_n)\!=\!i_n].\nonumber
\eqn
Introducing the time-dependent measure $\mu^t$ with $\mu^t(\{i\})\!:=\!\mu^{t}_{i}\!=\!\mathbb{P}[X(t)\!=\!i]$, we can express this as
\beqn
\mu_{i}^t=\sum_{1\leq j\leq k}M_{ij}(t,s)\mu_{j}^s,\nonumber
\eqn
for all $s<t$, and for $M_{ij}(t,s)\!:=\mathbb{P}[X(t)\!=\!i|X(s)\!=\!j]$. If we consider the $\left(M_{ij}(t,s)\right)_{1\leq i,j\leq k}$ as the matrix elements of a linear operator $M(t,s)$ acting on the vector space $\mathbb{R}^k\supset \mathcal{M}(K)$ with basis elements $\delta_1,\delta_2,\ldots,\delta_k$, we see that, as promised, for a Markov process the time evolution of the probability distribution is given by a linear map on $\mathbb{R}^k$ defined by its action on time-dependent probability measures:
\beqn\label{eq:MusMutAbstract}
\mu^s&\stackrel{M(t,s)}{\mapsto}\mu^t,\\
\mu^t&=M(t,s)\mu^s.
\eqn
This linear map describes the general time-inhomogeneous finite state Markov process and can easily be extended to the whole of $\mathbb{R}^k$.

In \cite{sumner2005} JGS and PDJ considered stochastic matrices as elements of the general linear group, and used this property to study the structure of invariant polynomials (used as measures of entanglement in quantum physics) when evaluated on a phylogenetic tree. Presently we will define the Markov semigroup which serves to refine the definition of invariant functions to the more relevant case of a stochastic (but linear) time evolution.

Define the time-dependent \textit{rate matrix}, $Q(t)$, as a (continuous) one-parameter family of linear operators on the vector space $\mathcal{M}(K)$, which in the $\delta_1,\delta_2,\ldots,\delta_k$ basis has matrix elements satisfying:
\beqn
Q_{ij}(t)\geq 0,\quad \forall i\neq j;\qquad Q_{ii}(t)=-\sum_{j\neq i}Q_{ji}(t).\nonumber
\eqn
The summation conditions can be equivalently expressed by defining the vector $\theta=\delta_1+\delta_2+\ldots+\delta_k$ and its transpose $\theta^\top$, and setting
\beqn
\theta^\top Q(t)=0,\nonumber
\eqn
for all $t$.

The Markov semigroup on $k$ elements, $\mathfrak{M}(k)$, with parameters $0\leq s\leq t< \infty$, is defined as the subset of (differentiable) two-parameter linear operators on $\mathcal{M}(K)$ which satisfy
\beqn
M(t,s)=1, \qquad \forall t=s;\nonumber
\eqn
the Chapman-Kolmogorov equation:
\beqn
M(t,s)M(s,r)=M(t,r),\qquad\forall r< s;\nonumber
\eqn
and the backwards and forwards equations:
\beqn\label{eq:Kolmogorov}
\frac{\partial M(t,s)}{\partial s}&=-M(t,s)Q(s),\\
\frac{\partial M(t,s)}{\partial t}&=Q(t)M(t,s);
\eqn
for any rate matrix $Q(t)$ \cite{goodman1970,isoifescu1980}.
Solutions of (\ref{eq:Kolmogorov}) can be represented using the time-ordered product (or ordered-exponential):
\beqn\label{eq:timeOrdered}
M(t,s)=\mathbb{T}\exp{\int_s^tQ(u)du}
\eqn
\cite[Chap. 4]{itzykson1980}, from which it follows that
\beqn\label{eq:markovDet}
\det{M(t,s)}=\exp{\int_s^ttr(Q(u))du},
\eqn
and
\beqn
\theta^\top M(t,s)=\theta^\top.\nonumber
\eqn
The time-ordered product is best understood by considering the approximation
\beqn
M(s+2\epsilon,s)=M(s+2\epsilon,s+\epsilon)M(s+\epsilon,s)\simeq e^{Q(s+\epsilon)\epsilon}e^{Q(s)\epsilon}.\nonumber
\eqn

By considering (\ref{eq:Kolmogorov}) for the case $t\!=\!s$, it follows that in the $\delta_1,\delta_2,\ldots,\delta_k$ basis, the matrix elements of each $M(t,s)$ lie in the interval $[0,1]$ for all $s\leq t$. Thus, the Markov semigroup corresponds to the subset of the set of stochastic matrices subject to the condition that for each matrix there exists a rate matrix (or generator) $Q(t)$ such that (\ref{eq:timeOrdered}) is satisfied. We refer to elements of the Markov semigroup as \emph{Markov operators}.

In the time-homogeneous case where the rate matrix is time-independent: 
\beqn
Q:=Q(t)=Q(0),\nonumber
\eqn 
it follows that $M(t,s)$ is dependent only upon the difference $(t-s)$, and form (\ref{eq:timeOrdered}) becomes simply
\beqn
M(t)=e^{tQ}=\sum_{0\leq n<\infty}\frac{(tQ)^n}{n!}.\nonumber
\eqn
In \S\ref{sec:Groups} we will discuss some representation-theoretic properties of certain groups affiliated with the Markov semigroup.

\subsection{Phylogenetic tensors}\label{sec:phylotens}

A tree, $\mathcal{T}$, is a connected graph without cycles and consists of a set of vertices and edges. Vertices of degree one are called \textit{leaves}. We work with \emph{oriented} trees, which are defined by directing each edge of $\mathcal{T}$ away from a distinguished vertex, $\rho$, known as the \textit{root} of the tree.
Consequently, a given edge lying between adjacent vertices $u$ and $v$ is specified as an ordered pair $(u,v)$, where $u$ lies on the unique path from $\rho$ to $v$. A \emph{cherry} is a pair of leaf vertices with the same parent vertex.

Assign a random variable, $X_v$, to each vertex of the tree, and, as described in \cite[Chap. 8]{semple2003},  a joint distribution of the random variables at the leaves is determined by specifying a distribution $\pi\!\in\!\mathcal{M}(K)$ at $\rho$ and a Markov operator $M^{v,u}\!\in\!\mathfrak{M}(k)$ for every edge $(u,v)$. In particular, for every $v$, the random variable $X_v$ is conditional on only the random variables lying on the path from $\rho$ and $v$, and for each pair of vertices $v_1,v_2$ with common parent $u$, the joint distribution of $X_{v_1}$ and $X_{v_2}$ is given by
\beqn\label{eq:branchprocess}
\mathbb{P}[X_{v_1}\!=\!i_1,X_{v_2}\!=\!i_2]=\sum_{1\leq j\leq k}M^{v_1,u}_{i_1j}M^{v_2,u}_{i_2j}\mu_j^u,
\eqn
where $\mu^u$ is the distribution of $X_u$. The empirical interpretation of the joint distribution across the leaves is that of a sampling distribution from which an alignment of molecular sequences is constructed by drawing one character pattern at a time. 
Throughout this paper, we will consider phylogenetic trees where the root distribution and the Markov operators are arbitrary.

Note that we insist that the Markov operators belong to the Markov semigroup, so a continuous-time process is in action throughout the tree.
This additional analytic structure means that this model is slightly less general than the \emph{general Markov model} as defined in \cite{allman2003,jayaswal2005}.
The general Markov model allows for arbitrary transition matrices with positive entries and unit row-sum (unit column-sum in our formulation), and it is not hard to find a matrix satisfying these conditions but with determinant less than or equal to zero, directly contradicting (\ref{eq:markovDet}).

We will consider a joint probability distribution on $m$ leaves as a probability measure, $P\!\in\otimes^m\mathcal{M}(K)$, that we refer to as a \textit{phylogenetic tensor}. Presently we review how these tensors can be constructed using purely algebraic operations.

\begin{figure}[tbp]
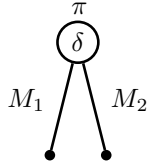

    \centering
    \rput[br](0.5,0.75){$\pi$}
    \pstree{\Tcircle{$\delta$}}{
        \Tdot \tlput{$M_{1}$}
        \Tdot \trput{$M_{2}$}
    }
    \caption{Phylogenetic tree with two leaves}
    \label{fig:2leaf}
\end{figure}

The branching process (\ref{eq:branchprocess}) can be interpreted as a map that takes probability measures on $K$ to probability measures on $K\!\times\!K\!=\!K^2$. In \cite{jarvis2005,sumner2005}, it was shown how to formalize this by defining the linear operator 
\beqn
\delta:\mathcal{M}(K)\rightarrow \mathcal{M}(K)\otimes\mathcal{M}(K).\nonumber
\eqn 
Demanding the conditional dependencies that are required for the standard definition of a tree distribution \cite[Chap. 8]{semple2003}, we have (expressed in the $\delta_1,\delta_2,\ldots, \delta_k$ basis) the specification
\beqn
\delta:\delta_i\mapsto\delta_i\otimes \delta_i,\qquad 1\leq i\leq k.\nonumber
\eqn
The phylogenetic tree with two leaves (Figure~\ref{fig:2leaf}) can then be represented as the string
\beqn
P=(M_1\otimes M_2)\cdot (\delta\cdot\pi),\nonumber
\eqn
where $M_1$ and $M_2$ are the Markov operators on the two edges of the tree and, if $X_1$ and $X_2$ are the random variables at the leaves 1 and 2, respectively, we have
\beqn
\mathbb{P}[X_1\!=\!i,X_2\!=\!j]=P_{ij}:=P(\{i\}\!\times\!\{j\}).\nonumber
\eqn

This construction can be generalized to any phylogenetic tree by colouring the root of the tree with a distribution $\pi$, each internal vertex (including the root) with a branching operator $\delta$, and every edge with an arbitrary Markov operator. The phylogenetic tensor is constructed by beginning at the root of the tree, and then recursively moving to the child vertices and applying the relevant operators to the corresponding slots in the (growing) tensor. 
Whenever a leaf is encountered, continually apply the identity operator at that leaf, until all leaves have been reached and the phylogenetic tensor is complete.
A phylogenetic tensor, $P$, is then represented as a string made up of the characters $\pi$, $M_1,M_2,\ldots,$ and $\delta$, and the joint distribution of the random variables $X_1,X_2,\ldots, X_m$ at the leaves $1,2,\ldots,m$ is given by
\beqn
\mathbb{P}[X_1\!=\!i_1,X_2\!=\!i_2,\ldots, X_m\!=\!i_m]=P_{i_1i_2\ldots i_m}:=\!P(\{i_1\}\!\times\!\{i_2\}\!\times\!\ldots\times\!\{i_m\}).\nonumber
\eqn

For example, the phylogenetic tensor of four leaves (Figure~\ref{fig:4leafConstruction}) is represented by the string
\beqn
P= (1\otimes 1\otimes M_3\otimes M_4)\cdot (1\otimes 1\otimes \delta)\cdot(1\otimes M_2\otimes M_5)\cdot(1\otimes \delta)\cdot (M_1\otimes M_6)\cdot(\delta\cdot\pi),\nonumber
\eqn
and is constructed in the steps 
\beqn
\pi
&\rightarrow  \delta\cdot\pi
\rightarrow (M_1\otimes M_6)\cdot(\delta\cdot\pi)
\rightarrow (1\otimes \delta)\cdot (M_1\otimes M_6)\cdot(\delta\cdot\pi)\nonumber\\
&\hspace{1em}\rightarrow (1\otimes M_2\otimes M_5)\cdot(1\otimes \delta)\cdot (M_1\otimes M_6)\cdot(\delta\cdot\pi)\\
&\hspace{2em}\rightarrow (1\otimes 1\otimes \delta)\cdot(1\otimes M_2\otimes M_5)\cdot(1\otimes \delta)\cdot (M_1\otimes M_6)\cdot(\delta\cdot\pi)\\
&\hspace{3em}\rightarrow (1\otimes 1\otimes M_3\otimes M_4)\cdot (1\otimes 1\otimes \delta)\cdot(1\otimes M_2\otimes M_5)\cdot(1\otimes \delta)\cdot (M_1\otimes M_6)\cdot(\delta\cdot\pi).
\eqn
In order to define Markov invariants, we must also define two reduced tensors based on $P$, the \emph{trimmed} tensor $\widetilde{P}$ and the \emph{pruned} tensor $P^\ast$.  These are both constructed by modifying the underlying tree.  The trimmed tensor $\widetilde{P}$ is constructed by taking $P$ and setting the Markov operators on the pendant edges all equal to the identity operator, or equivalently setting the lengths of the pendant edges to zero.  The pruned tensor $P^\ast$ is constructed by removing all cherries from the trimmed tensor. The rank of the pruned tensor is $(m-c)$ where $c$ is the number of cherries on the underlying tree.
\input{bigfig}

In the general case, we can relate $P$ and $\widetilde{P}$ as
\beqn\label{eq:PhylogeneticTreeModel}
P=(M_1\otimes M_2\otimes\ldots\otimes M_m)\cdot\widetilde{P},
\eqn
where $M_1,M_2,\ldots,M_m$ are the Markov operators on the leaf edges. In what is to come, we will continually use this relation.

As an illustration of the relationship between $P$, $\widetilde{P}$ and $P^\ast$, take the seven leaf tree (Figure~\ref{fig:7leaf}), with phylogenetic tensor given by
\beqn
P=(1\otimes M_2\otimes M_3\otimes M_4 &\otimes M_5\otimes M_6\otimes M_7)\cdot (1\otimes\delta\otimes\delta\otimes\delta)\\
&\cdot (M_1\otimes M_8\otimes M_9\otimes M_{10})\cdot (\delta\otimes\delta)\cdot (M_{11}\otimes M_{12})\cdot (\delta\cdot \pi).\nonumber
\eqn
The trimmed tensor corresponding to the tree (Figure~\ref{fig:trimmed7leaf}) is obtained by clipping off the pendant edges:
\beqn
\widetilde{P}=(1\otimes\delta\otimes\delta\otimes\delta)\cdot (1\otimes M_8\otimes M_9\otimes M_{10})\cdot (\delta\otimes\delta)\cdot (M_{11}\otimes M_{12})\cdot (\delta\cdot \pi),\nonumber
\eqn
and, finally the pruned tensor corresponding to the tree (Figure~\ref{fig:pruned7leaf}) is expressed as:
\beqn
P^\ast=(M_8\otimes M_9\otimes M_{10})\cdot (1\otimes\delta) \cdot (M_{11}\otimes M_{12})\cdot (\delta\cdot \pi).\nonumber 
\eqn
\input{7leaf}
\input{trimmed7leaf}
\input{pruned7leaf}

\subsection{Markov invariants, definition}\label{subsec:MarkovInvariantsDef}

With the form (\ref{eq:PhylogeneticTreeModel}) in mind, we define a \emph{Markov invariant} of \emph{weight} $(w_1,w_2,\ldots,w_m)$ as a function satisfying
\beqn\label{eq:markovinvdef}
f(P)=(\det M_1)^{w_1}(\det M_2)^{w_2}\ldots (\det M_m)^{w_m}f(\widetilde{P}),
\eqn
for all $M_1,M_2,\ldots,M_m\in\mathfrak{M}(k)$. 
We exclusively consider polynomial functions, and where $w_1\!=\!w_2\!=\ldots=\!w_m\equiv w$, the Markov invariant is said to be of weight $w$. 

Considering the above discussion of unbiased estimators of random variables, an unbiased estimator of a Markov invariant is a function, $\widehat{f}$, such that
\beqn
E[\widehat{f}(Z)]=f(P)=(\det M_1)^{w_1}(\det M_2)^{w_2}\ldots (\det M_m)^{w_m}f(\widetilde{P}).\nonumber
\eqn
Such an estimator depends, up to the multiplicative scaling factor, only upon the \textit{internal} structure of the phylogenetic tree. It is exactly this property that can be productively engaged in the context of phylogenetic tree inference.

Conversely, a \emph{phylogenetic invariant} is a function satisfying $f(P)\equiv 0$ for all $P$ belonging to the family of phylogenetic tensors arising from a particular tree (or subset of trees). In \S\ref{sec:markovinv} we will show that there exist Markov invariants for trees with three and four leaves that are simultaneously phylogenetic invariants.

Given a Markov invariant, $f$, consider the induced function, $f^\ast$,  defined on pruned tensors and specified by evaluating the trimmed tensor:
\beqn
f^\ast\!({P}^\ast)=f(\widetilde{P}).\nonumber
\eqn
This induced function is easily extended to be defined upon all of $\otimes^{m-c}\mathcal{M}(K)$, where $c$ is the number of cherries on the underlying tree of $P$. Such cases are of special interest for phylogenetic problems. In \S\ref{sec:markovinv} we will review a case (reported in a less general context in \cite{sumner2006}) where this induced function is itself a Markov invariant. We expect that future investigations of Markov invariants will reveal more cases such as this.

In \S \ref{subsec:theorem} we will establish existence conditions for Markov invariants using standard results from group representation theory.

%% file: bigfig.tex
\begin{figure}[tbp]
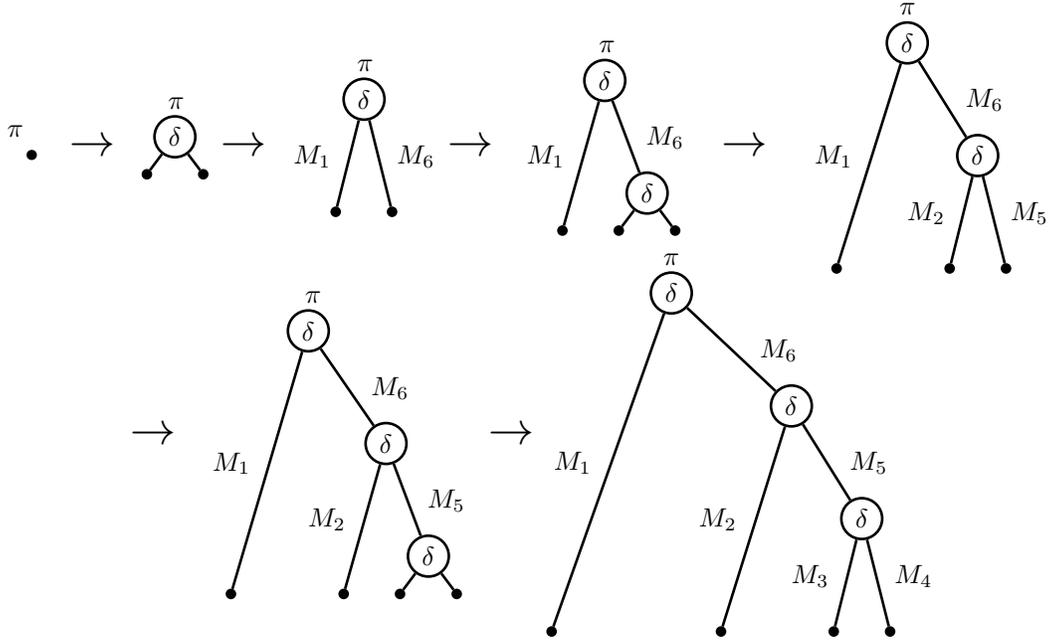

	\centering
	\begin{tabular}{ccccc}
		
		\raisebox{-15mm}{\rput[br](0.125,0.5){$\pi$}~
		\rput(0,0){$\bullet$} \hspace{5mm}\nextstep} 
		& 
		\raisebox{-12.5mm}{\rput[br](0.5,0.75){$\pi$}
			\pstree[levelsep=5mm]{\Tcircle{$\delta$}}{%
				\zerosplit
			}
		}
		\raisebox{-15mm}{\nextstep\hspace{5mm}} 
		& 
		\raisebox{-7.5mm}{\rput[br](0.5,0.75){$\pi$}
			\pstree[levelsep=15mm]{\Tcircle{$\delta$}}{%
				\Tdot \tlput{$M_{1}$} 
				\Tdot \trput{$M_{6}$}
			}}~
		\raisebox{-15mm}{\hspace{5mm}\nextstep\hspace{5mm}} 
		& 
		\raisebox{-5mm}{\rput[br](0.7,0.75){$\pi$}
			\pstree[levelsep=5mm]{\Tcircle{$\delta$}}{
				\skiplevels{3}
					\Tdot \tlput{$M_{1}$}
				\endskiplevels
				\skiplevels{2}
				\pstree[levelsep=5mm]{\Tcircle{$\delta$} \trput{$M_{6}$}}{
					\zerosplit
				}
				\endskiplevels
			}}
		\raisebox{-15mm}{\hspace{5mm}\nextstep\hspace{5mm}} 
		&
		\rput[br](1.05,0.75){$\pi$}
		\pstree[levelsep=5mm]{\Tcircle{$\delta$}}{
			\skiplevels{5}
				\Tdot \tlput{$M_{1}$}
			\endskiplevels
			\Tn
			\skiplevels{2}
			\pstree[levelsep=5mm]{\Tcircle{$\delta$} \trput{$M_{6}$}}{
				\skiplevels{2}
					\Tdot \tlput{$M_{2}$}
					\Tdot \trput{$M_{5}$}
				\endskiplevels
			}
			\endskiplevels
		}
	\end{tabular}
	
	\begin{tabular}{cc}
		\raisebox{-20mm}{\nextstep\hspace{5mm}}~
		\raisebox{-5mm}{\rput[br](1.2,0.75){$\pi$}
		\pstree[levelsep=5mm]{\Tcircle{$\delta$}}{
			\skiplevels{6}
				\Tdot \tlput{$M_{1}$}
			\endskiplevels
			\Tn
			\skiplevels{2}
			\pstree{\Tcircle{$\delta$} \trput{$M_{6}$}}{
				\skiplevels{3}
					\Tdot \tlput{$M_{2}$}
				\endskiplevels
				\skiplevels{2}
					\pstree{\Tcircle{$\delta$} \trput{$M_{5}$}}{
						\zerosplit
					}
				\endskiplevels
			}
			\endskiplevels
		}}
	& 
	\raisebox{-20mm}{\nextstep}~
	\rput[br](1.7,0.75){$\pi$}
	\pstree[levelsep=5mm]{\Tcircle{$\delta$}}{
		\skiplevels{8}
			\Tdot \tlput{$M_{1}$}
		\endskiplevels
		\Tn \Tn
		\skiplevels{2}
		\pstree{\Tcircle{$\delta$} \trput{$M_{6}$}}{
			\skiplevels{5}
				\Tdot \tlput{$M_{2}$}
			\endskiplevels
			\Tn
			\skiplevels{2}
				\pstree{\Tcircle{$\delta$} \trput{$M_{5}$}}{
					\skiplevels{2}
						\Tdot \tlput{$M_{3}$}
						\Tdot \trput{$M_{4}$}
					\endskiplevels
				}
			\endskiplevels
		}
		\endskiplevels
	}
	\end{tabular}
	
	\caption{Constructing the phylogenetic tensor for a four taxon tree}
	\label{fig:4leafConstruction}
\end{figure}

%% file: 7leaf.tex
\begin{figure}[b]
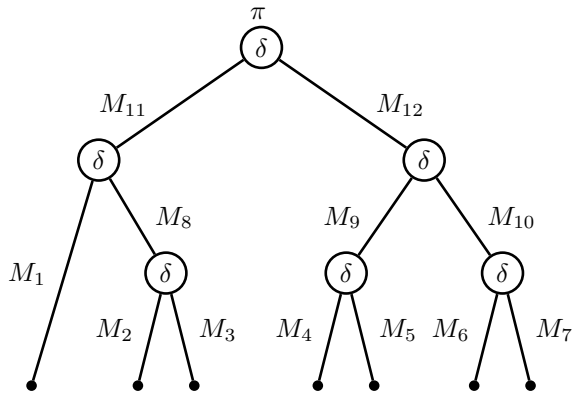

    \centering
    \rput[br](3.125,0.75){$\pi$}
    \pstree{\Tcircle{$\delta$}}{%
        \pstree{\Tcircle{$\delta$} \tlput{$M_{11}$}}{
            \skiplevel{\Tdot \tlput{$M_{1}$}}
            \Tn
            \pstree{\Tcircle{$\delta$} \trput{$M_{8}$}}{
                \Tdot \tlput{$M_{2}$}
                \Tdot \trput{$M_{3}$}
            }
        }
        \Tn \Tn \Tn \Tn
        \pstree{\Tcircle{$\delta$} \trput{$M_{12}$}}{
            \pstree{\Tcircle{$\delta$} \tlput{$M_{9}$}}{
                \Tdot \tlput{$M_{4}$}
                \Tdot \trput{$M_{5}$}
            }
            \Tn
            \pstree{\Tcircle{$\delta$} \trput{$M_{10}$}}{
                \Tdot \tlput{$M_{6}$}
                \Tdot \trput{$M_{7}$}
            }
        }
    }
    \caption{Phylogenetic tensor $P$ for seven taxon tree ((1,23),(45,67))}
    \label{fig:7leaf}
\end{figure}

%% file: trimmed7leaf.tex
\begin{figure}[tbp]
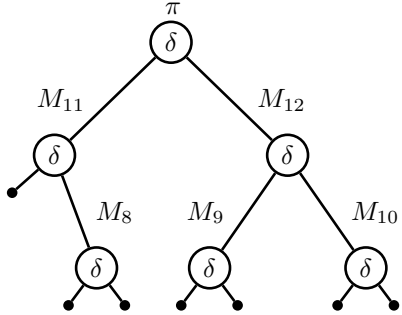

    \centering
    \rput[br](2.25,0.75){$\pi$}
    \pstree{\Tcircle{$\delta$}}{%
        \pstree[levelsep=5mm]{\Tcircle{$\delta$} \tlput{$M_{11}$}}{
            \Tdot
            \skiplevels{2}
                \pstree[levelsep=5mm]{\Tcircle{$\delta$} \trput{$M_{8}$}}{
                    \Tdot
                    \Tdot
                }
            \endskiplevels
        }
        \Tn \Tn
        \pstree{\Tcircle{$\delta$} \trput{$M_{12}$}}{
            \pstree[levelsep=5mm]{\Tcircle{$\delta$} \tlput{$M_{9}$}}{
                \Tdot
                \Tdot
            }
            \Tn
            \pstree[levelsep=5mm]{\Tcircle{$\delta$} \trput{$M_{10}$}}{
                \Tdot
                \Tdot
            }
        }
    }
    \caption{Trimmed phylogenetic tensor $\widetilde{P}$}
    \label{fig:trimmed7leaf}
\end{figure}

%% file: pruned7leaf.tex
\begin{figure}[tbp]
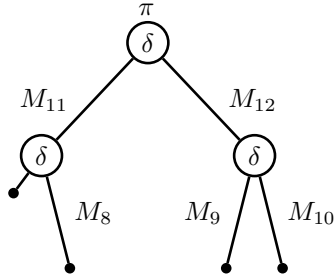

    \centering
    \rput[br](1.875,0.75){$\pi$}
    \pstree{\Tcircle{$\delta$}}{%
        \pstree[levelsep=5mm]{\Tcircle{$\delta$} \tlput{$M_{11}$}}{
            \Tdot
            \skiplevels{2}
				\Tdot \trput{$M_{8}$}
			\endskiplevels
        }
        \Tn \Tn 
        \pstree{\Tcircle{$\delta$} \trput{$M_{12}$}}{
            \Tdot \tlput{$M_{9}$}
            \Tdot \trput{$M_{10}$}
        }
    }
    \caption{Pruned phylogenetic tensor $P^\ast$}
    \label{fig:pruned7leaf}
\end{figure}

%% file: sec3.tex
\section{Group representation theory in phylogenetics}
\label{sec:Groups}
In this section we use the algebraic description of probability distributions on phylogenetic trees given in \S\ref{sec:phylotens} to establish natural connections with aspects of representation theory, for certain groups affiliated to the Markov semigroup. These are discussed in \S \ref{subsec:GroupDefns}. Then follows (\S \ref{subsec:RepGLK} and \S\ref{subsec:mtimesglk}) a brief outline of those aspects of the representation theory of the general linear group and its subgroups that are needed for the discussion of group branching rules. This leads to the construction of one-dimensional representations and their identification as invariants (\S \ref{subsec:InvariantsPlethysms}), with existence conditions given in \S\ref{subsec:theorem}. 

\subsection{The Markov semigroup and affiliated groups}
\label{subsec:GroupDefns}

The linear transformation (\ref{eq:MusMutAbstract}) effected under the Markov semigroup on probability measures is closely related to certain group actions on the vector space ${\mathbb R}^k$. Given that the corresponding representation theory is unchanged \cite{keown1975}, in this section we will generalize to complex vector space, (as with \cite{allman2003}). That is, here and below, for algebraic purposes we regard the $\delta_1,\delta_2,\ldots,\delta_k$ as elements of a basis for $V\cong {\mathbb C}^k$. Thus, the probability measures become a subset lying in the ambient complex space $\otimes^m {\mathbb C}^k \supset \otimes^m\mathcal{M}(K) $.
For related considerations involving the study of invariants of stochastic matrices see \cite{johnson1985,mourad2004}. 

Referring to (\ref{eq:markovDet}) and noting that $-\infty < tr(Q(t))\leq 0$ for all $t$, the determinant of each element $M(t,s)$ lies in the interval $(0,1]$, and the Markov semigroup occurs as a subset of the general linear group:
\beqn
\mathfrak{M}(k)\subset GL(k).\nonumber
\eqn
$GL(k)$ is the group of \textit{invertible} linear operators on the $k$-dimensional vector space $\mathbb{C}^k$. The smallest subgroup of $GL(k)$ that contains $\mathfrak{M}(k)$ is obtained by taking $\mathfrak{M}(k)$ together with all of its operator inverses. In order to apply known methods of representation theory, we will, however, not work with this group directly. We define a slightly less refined subgroup as the focus of the impending discussion.

Generalizing the notation of \cite{mourad2004}, we define the subgroup $GL_1(k)\lhd GL(k)$ as the subset of $GL(k)$ whose matrices in the $\delta_1,\delta_2,\ldots,\delta_k$ basis have unit column-sum. That is, for all $g\in GL_1(k)$:
\beqn\label{eq:gl1k}
\theta^\top g=\theta^\top.\nonumber
\eqn
The group property clearly holds, as for all $g_1,g_2\in GL_1(k)$:
\beqn
\theta^\top(g_1g_2)=(\theta^\top g_1)g_2=\theta^\top g_2=\theta^\top.\nonumber
\eqn
This group is isomorphic to the complex \emph{affine group}\footnote{The symbol $\ltimes$ denotes the \textit{semi-direct product} of two groups \cite[Chap. 1]{baker2003}. The standard physical example is the Euclidean group, which occurs as the semi-direct product between rotations and translations in $\mathbb{R}^n$. These are none other than the set of transformations that define Euclidean geometry.} 
\[
GL(k\!-\!1) \ltimes T(k-1) \equiv A(k-1),
\]
where $T(k-1)$ is the group of linear translations on $\mathbb{C}^{k-1}$. As shown in Appendix~\ref{sec:Proofs}, this isomorphism is due to the column-sum condition being, in effect, a statement that the group elements are dual to linear transformations in $k$-dimensional complex space, leaving a fixed vector invariant.

Consider also the \emph{doubly-stochastic} Markov semigroup, $\mathfrak{M}^\ast\!(k)$, obtained by requiring an additional condition on the rate matrices:
\beqn
Q(t)\theta=0.\nonumber
\eqn
The associated subgroup of the general linear group is then denoted as $GL_{1,1}(k)$; the subgroup of matrices in $GL(k)$ which have unit column- \textit{and} row-sum with, for all  $g\in GL_{1,1}(k)$:
\beqn
\theta^\top g&=\theta^\top,\\
g\theta&=\theta.\nonumber
\eqn
Again the group property can easily be shown to hold. Thus the doubly-stochastic Markov semigroup is naturally affiliated to the associated group $GL_{1,1}(k)$ which, also as shown in Appendix~\ref{sec:Proofs}, itself is isomorphic to $GL(k\!-\!1)$.

To summarise, consider the subgroup chain:
\beqn\label{eq:subgroupchains}
GL(k\!-\!1)\cong GL_{1,1}(k)\lhd GL(k\!-\!1) \ltimes T(k\!-\!1)\equiv A(k)\cong GL_1(k)\lhd GL(k). 
\eqn
and the set inclusions:
\beqn
\mathfrak{M}(k)&\subset GL_1(k),\nonumber\\
\mathfrak{M}^\ast\!(k)&\subset GL_{1,1}(k).
\eqn

We now have a clear picture of how to develop the representation theory of the Markov semigroup which focuses on algebraic properties and avoids the analytic details due to the positivity requirement and semigroup property. This is the correct framework in which to exploit the Schur-Weyl duality (\S\ref{subsec:RepGLK}) and, considering the above inclusions, all results presented will be valid for the Markov semigroup. The above subgroup chain will feature in \S\ref{subsec:InvariantsPlethysms} where we give existence conditions for Markov invariants.

\subsection{Representations of $GL(k)$ and Schur-Weyl duality}
\label{subsec:RepGLK}

Our purpose here is to show that the close relation of the Markov semigroup to affiliated subgroups of the general linear group allows the machinery of representation theory to be applied in analysing the models used in phylogenetic inference.

From the form of the general Markov model on phylogenetic trees given earlier (\ref{eq:PhylogeneticTreeModel}), it is evident that the representation-theoretic considerations must be extended to tensor products. 
We now provide some standard results within this setting (see, for example, \cite[Lecture 6]{fulton1991}).

For $GL(k)$ and its classical subgroups it is well known that for the \emph{defining} representation on $V \cong {\mathbb C}^k$, with $v \mapsto gv$, extended to a reducible representation on $\otimes^m V$ in the obvious way, $v_1 \otimes v_2 \otimes \ldots \otimes v_m \mapsto gv_1 \otimes gv_2 \otimes \ldots\otimes gv_m$, there is a direct sum decomposition,
\begin{align}
\label{eq:SchurWeylDuality}
\otimes^m V = & \, \sum_{\lambda\vdash m}\oplus f_\lambda V^\lambda ,
\end{align}
into (possibly reducible) subspaces $V^\lambda$. These subspaces (or \textit{modules}) are labelled by integer partitions, $\lambda= (\lambda_1, \lambda_2, \ldots, \lambda_n)$, of $m$, the $\lambda_i$ being nonzero and nonincreasing and such that $\lambda_1 + \lambda_2 + \ldots + \lambda_n = m$. If $\lambda$ is a partition of $m$, we write $\lambda \vdash m$ and $|\lambda|\!=\!m$. The corresponding module $V^\lambda$ is determined by a unique projector on $\otimes^m V$; the \textit{Young's operator} $Y^\lambda$. The $f_\lambda$ are integer multiplicities determining how many times each module occurs in the decomposition. The \emph{Schur-Weyl duality} is the classic result that each $f_\lambda$ is none other than the dimension of the irreducible representation associated with the same partition $\lambda$ of the \text{symmetric} group ${\mathfrak S}_m$. This reflects the role of the symmetric group's action on $\otimes^m V$ by permuting basis elements across the tensor product, when constructing the Young's operators.

The \textit{character} of a representation is defined as the set of traces of the representing matrices; one for each group element. The irreducible representations of a group can be enumerated by solely considering the corresponding irreducible characters. Thus the problem of decomposing a representation into irreducible modules (computing the multiplicities $f_\lambda$) can be performed at the level of the characters\footnote{Within the context of phylogenetics, see \cite{matsen2006} for an unrelated discussion of the irreducible characters of the symmetric group.}.

For instance, in the case of $GL(k)$ itself, the $V^\lambda$ are irreducible, with character given by the celebrated Schur functions, $s_\lambda$, with 
\[
s_\lambda(x) = \mbox{tr}(\pi_\lambda(g)),
\]
where $\pi_\lambda(g)$ is the representing matrix for group element $g$ and $x_1, x_2, \ldots,x_k$ are its eigenvalues.  The Schur functions are defined in their own right, and are uniquely determined by the semi-standard tableaux corresponding to the partition $\lambda$ \cite{macdonald1979}.

The defining $k$-dimensional representation in this notation is $V^{\{1\}} \cong {\mathbb C}^k$, in which case the Schur function is $s_{\{1\}}(x)=x_1 + x_2 + \ldots + x_k$. The Schur functions form a basis for the ring of symmetric functions on any number of variables, and the trace is a symmetric function. Hence, the problem of identifying the irreducible $GL(k)$ modules in the above representation on $\otimes^mV$, reduces to identifying the Schur functions in the decomposition of the character with respect to this basis\footnote{The stronger statement that the $V^\lambda$ provide the complete set of irreducible modules of \emph{any} integral representation of $GL(k)$ is valid \cite{keown1975}.}.

A convenient and standard notation for Schur functions is given by enclosing the partition (or parts thereof) in braces \cite{littlewood1940}. Thus $\{\lambda\}$ and $\{1\}$, are the Schur functions corresponding to a general irreducible and the defining representation of $GL(k)$ respectively. For simplicity, we write $\pi_{\{1\}}(g)=g$.
 
For classical subgroups of $GL(k)$, the modules $V^\lambda$ are no longer necessarily irreducible, and further combinatorial considerations (not required here) are needed to effect a complete reduction\footnote{The classical subgroups of $GL(k)$ are constructed by requiring, under the group action, the invariance of bilinear forms on $V$.}. More importantly, for $GL(k)$ itself with $V$ not the defining, but an arbitrary module, $V^\rho$ say, the equivalents of the above modules, $(V^\rho)^\lambda$, are again no longer irreducible in general.

This construction introduces a fundamental operation for combining representations together; that of \emph{plethysm} \cite{littlewood1940}. The character of $(V^\rho)^\lambda$ is denoted $\{\rho\} \underline{\otimes} \{ \lambda \} $; the plethysm of $\{\rho\}$ by $\{ \lambda\}$. In the simplest case $\{\rho\}$ is the character for the  defining representation, $\{1\}$, and by definition $\{1\} \underline{\otimes}  \{\lambda\}  = \{\lambda\} $.

In general, for any symmetric functions $A,B$ we have $\{\rho\} \underline{\otimes} (A+B) = \{\rho\} \underline{\otimes} A + \{\rho\} \underline{\otimes} B$, and we recover
\begin{align}
\{\rho\} \underline{\otimes} \big( \sum_{\lambda \vdash m}  f_{\lambda } \{\lambda\} \big) 
= & \{\rho\}\otimes\{\rho\}\otimes\ldots\otimes\{\rho\}, \nonumber
\end{align}
where $\{\rho\}\otimes\{\lambda\}$ denotes the (commutative and associative) pointwise multiplication of the Schur functions, 
\beqn
(\{\rho\}\otimes \{\lambda\})(x):=\{\rho\}(x)\cdot \{\lambda\}(x)\nonumber,
\eqn
and the Schur functions occurring in the decomposition of $\{\rho\}\otimes \{\lambda\}$ correspond to partitions of $|\rho|+|\lambda|$. This of course reflects (\ref{eq:SchurWeylDuality}) with $V$ replaced by $V^\rho$:
\beqn
V^\rho\otimes V^\rho \otimes\ldots \otimes V^\rho=\sum_{\lambda \vdash m}\oplus f_\lambda {(V^\rho)}{}^\lambda\nonumber .
\eqn
In particular, for rank 2 we have
\begin{align}
V^\lambda\otimes V^\lambda=(V^\lambda)^{\{2\}}\oplus (V^\lambda)^{\{1^2\}},\nonumber
\end{align}
which at the level of the characters is described completely by
\begin{align}
\{\lambda\} \otimes \{\lambda\} = & \, \{\lambda\} \underline{\otimes} \{2\} \, + \, \{\lambda\} \underline{\otimes} \{1^2\}. \nonumber
\end{align}
This is the well-known decomposition of a representation into its symmetric and anti-symmetric Kronecker square, respectively\footnote{In the context of quantum physics, this corresponds exactly to the statistical properties for ensembles of bosonic and fermionic particles, respectively.}. 

Although it is a difficult task to evaluate the general plethysm (see \cite{macdonald1979} for a review of symmetric functions and their various manipulations), in practice all required operations of symmetric functions involving products, plethysms and group branching rules can be evaluated symbolically using an appropriate group theory package. Where required, we use \texttt{Schur} \cite{schur} for this purpose.

From (\ref{eq:PhylogeneticTreeModel}), which gives the form of the phylogenetic tensor for a tree with $m$ leaves, it is clear that the appropriate representation space to consider is indeed ${\otimes^m}{\mathbb C}^k$, regarded not as a module of $GL(k)$ as above, but rather carrying an irreducible representation of the action of the direct product group $\times^m GL(k) =GL(k) \times GL(k) \times \ldots \times GL(k)$. That is, considering that a phylogenetic tensor lies in the ambient space $\otimes^m\mathbb{C}^k$, the generic analogue of (\ref{eq:PhylogeneticTreeModel}) is
\beqn\label{eq:DirectProdRep}
\psi'=(g_1\otimes g_2\otimes \ldots \otimes g_m)\cdot\psi,
\eqn
where $\psi\in\otimes^m\mathbb{C}^k$. In a phylogenetic setting, we must allow for differing Markov operators to act on each edge; hence the above form. It is usual in phylogenetics to take a fixed rate matrix for all edges, and allow the edge lengths to vary, thus creating different Markov operators from identical generators. In fact, the above generalization of the group action allows for differing Markov \emph{processes} on every edge of the phylogenetic tree\footnote{Under a (somewhat biologically unsound) model in which the evolution along the pendant edges occurs with identical transition probabilities, the group becomes the diagonal $GL(k)$ subgroup of the $m$-fold direct product group, and the representation reduces accordingly, precisely as in the initial discussion above.}.

A complete representation-theoretic analysis incorporating the tree structure of phylogenetic tensors is a topic for future research, and we defer such a theory. We concentrate on analysing the group action defined by (\ref{eq:PhylogeneticTreeModel}), leading toward the derivation of Markov invariants while ignoring the underlying tree structure. In \S\ref{sec:markovinv} we will introduce a post-hoc procedure which allows the tree structure to be incorporated. This will allow Markov invariants to be applied in a practical setting without the need for the complete theory.

\subsection{Representations of $\times^mGL(k)$}\label{subsec:mtimesglk}

Here we derive the group branching rule which is required to identify the irreducible modules under the group action (\ref{eq:DirectProdRep}).

There is yet another product of symmetric functions; the \emph{inner product}, defined as
\beqn
\{\lambda\}\odot\{\rho\}=\sum_{\sigma\vdash n}\gamma^\sigma_{\lambda\rho}\{\sigma\},\nonumber
\eqn
where $|\lambda|\!=\!|\rho|\!=\!|\sigma|\!=n$, and the $\gamma^\sigma_{\lambda\rho}$ are the integer multiplicities of occurrences of the $\sigma$ representation in the Kronecker product representation between $\lambda$ and $\rho$ of the symmetric group $\mathfrak{S}_n$ \cite{littlewood1957}.

Consider the direct product group $GL(k)\times GL(\ell)$, with group action on $V_1\otimes V_2$, where $V_1$ is $k$-dimensional and $V_2$ is $\ell$-dimensional, defined by $v_1\otimes v_2\mapsto g_1v_1\otimes g_2v_2$. If the eigenvalues of $g_1,g_2$ are $x_1,x_2,\ldots,x_k$ and $y_1,y_2,\ldots,y_{\ell}$ respectively, then the character of this representation is the product
\beqn
{\{1\}}(x)\cdot {\{1\}}(y)=(x_1+\ldots +x_k)(y_1+\ldots +y_l)={\{1\}}(xy),\nonumber
\eqn
with
\beqn
(xy) = (x_1y_1, x_1y_2,\ldots, x_2y_1,\ldots, x_ky_\ell).\nonumber
\eqn 
Generalizing this result, consider the natural embedding, $GL(k)\times GL(\ell)\subset GL(k\ell)$, and the $\{\lambda\}$ representation of $GL(k\ell)$ restricted to the direct product group: $\Psi\mapsto \pi_\lambda (g_1\times g_2)\Psi$ with $\Psi\in (V_1\otimes V_2)^\lambda$. The character of this representation has decomposition
\beqn\label{eq:directpleth}
\{\lambda\}(xy)=\sum_{\rho,\sigma\vdash |\lambda|}\gamma^\lambda_{\rho\sigma}\{\rho\}(x)\cdot\{\sigma\}(y);
\eqn
for details see \cite{king1975,whippman1965}. Thus, we see that the inner product plays an essential role in decomposing representations of the direct product group $GL(k)\!\times\! GL(\ell)$ into tensor products of irreducible modules of $GL(k)$ with irreducible modules of $GL(\ell)$:
\beqn
(V_1\otimes V_2)^\lambda=\sum_{\rho,\sigma\vdash |\lambda|}\oplus\gamma^\lambda_{\rho\sigma}V_1^\rho\otimes V_2^\sigma.\nonumber
\eqn
Presently we will use this result to derive branching rules for the group action that is relevant to phylogenetics (\ref{eq:PhylogeneticTreeModel}).

In its general setting, a (group) \emph{branching rule} describes the decomposition of a representation of a group, $G$, when restricted to a subgroup, $H\subset G$ (written as $G\downarrow H$) \cite[Chap. V, \S 18]{weyl1950}. For the present purpose, we consider $\times^mGL(k)$ as a subgroup of $GL(k^m)$, and given the defining representation of $GL(k^m)$, the corresponding branching rule is
\beqn
GL(k^m)\downarrow \times^mGL(k):\quad\{1\}\longrightarrow \{1\}\otimes\{1\}\otimes\ldots\otimes \{1\}=\otimes^m\{1\}.\nonumber
\eqn
On the left-side of the arrow, $\{1\}$ denotes the defining representation of $GL(k^m)$, whereas on the right-side, $\{1\}$ denotes the defining representation of $GL(k)$.

If we take the generic $\{\lambda\}$ representation of $GL(k^m)$, the appropriate branching rule is\footnote{This is a special case of a more general embedding 
$\{1\} \rightarrow \{\lambda_1\} \otimes  \{\lambda_2\} \otimes \ldots \otimes  \{\lambda_m\}$, for which 
each $\{ \sigma_i\}$ in the decomposition is replaced by the appropriate plethysm
$\{\lambda_i \} \underline {\otimes} \{ \sigma_i \}$. For a recent discussion of the calculus of plethysms see \cite{fauser2006}.} 
\beqn
\label{eq:LeafBranchingRule}
&GL(k^m) \downarrow  \, \times^m GL(k):\quad\{ \lambda \} \longrightarrow   \, 
\sum
_{\sigma_1,\sigma_2,\ldots,\sigma_m \vdash |\lambda|}
^{ \{\sigma_1\} \odot \{\sigma_2\} \odot \ldots\odot\{\sigma_m\} \ni \{\lambda\}}
\{ \sigma_1 \} \otimes \{ \sigma _2 \} \otimes \ldots \otimes \{\sigma_m\}. 
\eqn
This result can be confirmed using the identity (\ref{eq:directpleth}).

The branching rule (\ref{eq:LeafBranchingRule}) gives the decomposition of irreducible modules of the time evolution at the pendant edges of a tree, as induced by (\ref{eq:PhylogeneticTreeModel}), but considered as the representation of $\times^m GL(k)\subset GL(k^m)$ defined by
\beqn
\Psi'=\pi_\lambda (g_1\times g_2\times\ldots\times g_m)\cdot\Psi,\nonumber
\eqn
for $\Psi\in(\otimes^m\mathbb{C}^k)^\lambda$.

In the setting of phylogenetics, we show in \S\ref{subsec:InvariantsPlethysms} that specializing to $\{\lambda\}\!\equiv\!\{d\}$ gives the decomposition of (homogeneous degree $d$) polynomials of phylogenetic tensors. In a practical setting, this corresponds exactly to taking (polynomial) transformations of the observed data set of character pattern counts. That is, recalling that the expectation value of character pattern counts in a sequence alignment is governed by a joint distribution on a tree corresponding to a phylogenetic tensor $P$, the above branching rule tells us how arbitrary polynomial functions of the character pattern counts decompose into components which transform among themselves under the time evolution given in (\ref{eq:PhylogeneticTreeModel}). 
In addition to what is presented here, this potentially has application to any analysis involving pattern counts in molecular sequence data (see \S\ref{sec:Outlook} for further comments on this matter).
 
In \S\ref{subsec:theorem}, we will exploit the branching rule directly, defining the one-dimensional modules in the decomposition  (\ref{eq:LeafBranchingRule})  as \emph{invariants}, and give existence conditions for Markov invariants. We must first establish the isomorphism between homogeneous degree $d$ polynomials on a vector space $V$, and the module $V^{\{d\}}$.

\subsection{Symmetric plethysms and invariants}
\label{subsec:InvariantsPlethysms}
Associated with any representation $V$ of a group $G$ is the so-called coordinate ring ${\mathcal P}{(}V{)}$ of polynomials\footnote{${\mathcal P}{(}V{)}$ is the ring of polynomials in the basis elements, $\xi_1,\xi_2,\ldots,\xi_k$, of the dual space $V^\ast$ so that ${\mathcal P}{(}V{)}\equiv \mathbb{C}[\xi_1,\xi_2,\ldots,\xi_k]$ with $\xi_i(\delta_j)=\delta_{ij}$ for all $1\leq i,j\leq k$.} over $\mathbb{C}$ in the components $v_1, v_2, \ldots ,v_k$, corresponding to a given basis for $V$. For such polynomials, $f(v)$, there is a natural group action,
\[
f(v) \rightarrow g\cdot f(v) := f(g^{-1} v).
\]
There is an isomorphism between the ring ${\mathcal P}{(}V{)}$ and the symmetric tensor algebra\footnote{See \cite[Chap. 4]{goodman1998} for a discussion of the symmetric tensor algebra.} $\mbox{\Large{$\vee$}}(V)$: 
\begin{align}
\label{eq:SymmetricAlgIso}
{\mathcal P}{(}V{)} \equiv & \, \sum_{d=0}^\infty {\mathcal P^d}{(}V{)}
\cong \mbox{\Large{$\vee$}}(V) \equiv \sum_{d=0}^\infty\mbox{\Large{$\vee$}}^d (V), 
\end{align}
with $\mbox{\Large{$\vee$}}^d (V)\cong V^{\{d\}}$ and ${\mathcal P^d}{(}V{)}$ denoting the homogeneous polynomials of degree $d$. This reflects that an arbitrary homogeneous polynomial of degree $d$ in $k$ indeterminates can be specified by an array of determinates $f_{i_1i_2\ldots i_{d}}$ which is symmetric under permutation of indices:
\beqn
f(v)=\sum_{1\leq i_1,i_2,\ldots ,i_d \leq k}f_{i_1i_2\ldots i_{d}}v_{i_1}v_{i_2}\ldots v_{i_d}.\nonumber
\eqn

Our interest in the above construction lies in the \emph{invariant ring}, ${\mathcal P}{(}V{)}{}^G$, of polynomials that are invariant up to a multiplicative factor under the action of $G$, or more generally, for any subgroup $H \unlhd G$,
\beqn\label{eq:invariantdef}
f(hv) = \det(h)^w f(v),
\eqn
for all $h\in H$ and $v\in V$. For matrix groups the multiplicative factor is the determinant with $w$ denoting the \emph{weight} of the invariant. Using the isomorphism (\ref{eq:SymmetricAlgIso}), the identification of a linear basis of such invariants of degree $d$ reduces to the identification, in the reduction of the $V^{\{d\}}$, of the one-dimensional representations of $H$ in the branching rule $G \downarrow H$.

In particular, the one-dimensional representations of $GL(k)$ occur as follows. Note that the dimension of a representation is equal to the trace of the representing matrix of the identity. For the irreducible module $V^{\lambda}$ this is given by $s_{\lambda}(1,1,\ldots ,1)$. Thus, for a one-dimensional module, the corresponding Schur function must be monomial and (considering the definition of the Schur functions using summations over semi-standard tableaux given in \cite{macdonald1979}) this occurs only for partitions of the form $\{r^k\}$ for any integer $r\!>\!0$. Additionally, considering that one-dimensional representations act by simply multiplying by the character itself, and that
\beqn\label{eq:1Drep}
s_{\{r^k\}}(x)&=(x_1x_2\ldots x_k)^r,\\
&=\det(g)^r,
\eqn
we see that, for any $\psi\in V^{\{r^k\}}$, we have
\beqn
\psi\mapsto\det(g)^r\psi,\nonumber
\eqn
under the $\{r^k\}$ representation of $GL(k)$. This should be compared directly to (\ref{eq:invariantdef}).

Taking $\mathfrak{M}(k)\subset GL_1(k)$, we can construct Markov invariants by identifying polynomials lying in the invariant ring for $GL_1(k)$. Clearly, any polynomial
\[
f\in{\mathcal P}{(}\!\otimes^m V{)}^{\times^mGL_1(k)}
\]
must also satisfy (\ref{eq:markovinvdef}) and is hence a Markov invariant. Recalling the salient subgroup chain (\ref{eq:subgroupchains}), affiliated to the Markov semigroup, the representation-theoretic task is to evaluate the relevant branching rules for specific cases. The required branching rules derive from (\ref{eq:LeafBranchingRule}), together with identification of the specific form of one-dimensional representations of the subgroup in question.

It should be noted that this procedure leaves open the possibility that there exist Markov invariants that do not occur in the invariant ring for $GL_1(k)$. We leave this as an open problem, but note that it is plausible that such a possibility could be excluded by continuity arguments.  

\subsection{Markov invariants, existence theorems}\label{subsec:theorem}
Presently, we use the facts we have collected above to establish existence conditions for polynomial invariants for the group actions of $GL(k),GL_1(k)$ and $GL_{1,1}(k)$.

\medskip\noindent
\textbf{Theorem 1}: Polynomial invariants for phylogenetic models. \\
Linearly independent polynomial invariants at degree $d$ of the groups: 
\begin{enumerate}
\item[i.] $\times^m GL(k),$
\item[ii.] $\times^mGL_1(k),$ and
\item[iii.] $\times^m GL_{1,1}(k),$
\end{enumerate}
are given by the one-dimensional modules of these groups occurring in the decomposition of the $GL(k^m)$ module ${(\otimes^m V)}{}^{\{d\}}$. In each case the one-dimensional modules correspond to $m$-fold products of Schur functions labelled by partitions of $d$:
\begin{enumerate}
\item[i.] $\{r^k\} \otimes \{r^k\}\otimes \ldots \otimes \{r^k\},$ 
\item[ii.] $\{r_1\!+\!s_1,r_1^{k\!-\!1}\} \otimes \{r_2\!+\!s_2,r_2^{k\!-\!1}\}\otimes \ldots \otimes \{r_m\!+\!s_m,r_m^{k\!-\!1}\}$, and
\item[iii.] $\{r_1\!+\!s_1,r_1^{k\!-\!2},t_1\} \otimes \{r_2\!+\!s_2,r_2^{k\!-\!2},t_2\}\otimes \ldots \otimes \{r_m\!+\!s_m,r_m^{k\!-\!2},t_m\}$, respectively,
\end{enumerate}
with
\begin{enumerate}
\item[] $kr\equiv d$,
\item[] $kr_a+s_a\equiv d$, and
\item[] $(k-1)r_b+t_b+s_b\equiv d$,
\end{enumerate}
for all $1\leq a,b\leq m$ respectively. 

Given the isomorphism (\ref{eq:SymmetricAlgIso}) and the branching rule (\ref{eq:LeafBranchingRule}) with $\{\lambda\}\equiv \{d\}$, in each case the number of admissible partitions of the given forms $\{\sigma_1\} \otimes \{\sigma_1\} \otimes \ldots \otimes \{\sigma_m \}$ is the number of times the inner product $\{\sigma_1\} \odot\{\sigma_2\}\odot \ldots \odot\{ \sigma_m\}$ of irreducible representations of the symmetric group ${\mathfrak S}_d$ contains the one-dimensional irreducible representation $\{d\}$. This is also the number of linearly independent polynomial invariants in each case.

\medskip\noindent
\textbf{Proof}:
Each case identifies representations of $\times^m GL(k)$ with character $\{\sigma_1\} \otimes \{\sigma_2\} \otimes \ldots \otimes \{\sigma_m \}$, each component of which is a partition that corresponds to a one-dimensional representation of the respective subgroup. The dimension of this representation is the product of the dimension of each of the representations labelled by $\{\sigma_a\}$. Therefore the representation is one-dimensional if and only if, for each $\{\sigma_a\}$, the corresponding representation is one-dimensional.

For case (i), $GL(k)$, as we showed in \S\ref{subsec:InvariantsPlethysms}, the representation labelled by $\{r^k\}$ is one-dimensional, providing an invariant of weight $w\equiv r$. For case (ii), $GL_1(k)$, it is established in the appendix that the representation of $GL(k)$ labelled by $\{r_a\!+\!s_a,r_a^{k\!-\!1}\}$ contains a unique one-dimensional module under $GL_1(k)$. For case (iii), as will also be established in the appendix, $GL_{1,1}(k)$ is isomorphic to $GL(k\!-\!1)$ and the $GL(k)$ character $\{r_a\!+\!s_a,r_a^{k\!-\!2},t_a\}$ contains under branching to $GL(k\!-\!1)$, a unique one-dimensional module with character $\{ r_a^{k\!-\!1} \}$.

\hfill $\Box$

\noindent
Note that case (ii) is a special instance of case (iii), with $t_a\!=\!0$, and case (i) is a special instance of case (ii), with $s_a\!=\!0$. This reflects the definition (\ref{eq:invariantdef}).

Recall the inclusion
\beqn
\mathfrak{M}(k)\subset GL_1(k) \lhd GL(k).\nonumber
\eqn
It is clear that any invariant that exists for case (i), with $w\!\equiv \!r$, or (ii), with $w\!\equiv r_1\!=\!r_2\!=\!\ldots\!=\!r_m$, is necessarily a Markov invariant, (\ref{eq:markovinvdef}), with the particular form
\beqn\label{eq:markovinv}
f(P)=(\det M_1\det M_2\ldots \det M_m)^wf(\widetilde{P}).
\eqn 
In \S\ref{sec:markovinv} we will count occurrences of this type of Markov invariant for various cases of interest to phylogenetics; $k\!=\!2$ to 4 character states and trees with $m\!=\!2$ to 10 leaves. We will also briefly review the algebraic structure of these invariants in the cases $m\!=\!2$ to 4 when evaluated upon phylogenetic tensors, and give examples of how this structure can be gainfully employed in the problem of phylogenetic tree inference from molecular sequence data.

Taking case (ii) in its general form, we see that for $w_1\!\equiv r_1,w_2\!\equiv r_2,\ldots w_m\!\equiv r_m,$ it is possible that there exist Markov invariants, taking the general form
\beqn
f(P)=\left(\det M_1^{w_1}\det M_2^{w_2}\ldots \det M_m^{w_m}\right)f(\widetilde{P}).\nonumber
\eqn
When the distinction is required, we refer to these invariants as \emph{mixed weight Markov invariants}. In \S\ref{subsec:MixedWeight} we will show that such invariants do indeed exist in various cases of interest to phylogenetics. However, as yet the explicit form of these invariants has not been constructed, and their structure remains unexplored.

Recall the inclusion
\beqn
\mathfrak{M}^\ast\!(k)\subset GL_{1,1}(k) \subset GL(k),\nonumber
\eqn
for the doubly-stochastic Markov semigroup. The case (iii) establishes existence conditions for polynomial invariants for this semigroup. These invariants will be valid for any joint distribution on a phylogenetic tree which is constructed using only doubly-stochastic matrices. This includes oft-used models such as Jukes-Cantor, K80, K3ST and SYM \cite{zharkikh1994}. We report the above theorem, but defer the exploration of the invariants in this case.

%% file: sec4.tex
\section{Markov invariants in phylogenetics}\label{sec:markovinv}
In \S\ref{subsec:zoo} we  establish existence of Markov invariants relevant to phylogenetics for the cases of  $k\!=\!2$ to 4 character states, distinguishing between true Markov invariants and invariants which are valid for the full general linear group. In \S\ref{subsec:EvaluateTree} we report upon known algebraic relations between Markov invariants when evaluated upon phylogenetic tensors for $k\!=\!2$ to 4 character states and for trees with $m\!=\!2$ to 4 leaves. We also discuss the application of Markov invariants to the problem of phylogenetic tree reconstruction in these cases. Finally, in \S\ref{subsec:MixedWeight} we establish existence of mixed weight Markov invariants for $k\!=\!4$ character states and trees with $m\!=\!2$ to 5 leaves. Throughout, we used \texttt{Schur} \cite{schur} for all non-trivial manipulations of Schur functions.

\subsection{Zoo of invariants and nomenclature}\label{subsec:zoo}

We gave, in \S \ref{subsec:InvariantsPlethysms}, a sufficient condition for the existence of a Markov invariant, (\ref{eq:markovinv}), of degree $d$ and weight $w$:
\beqn\label{eq:SuffCond}
\{r+s,r^{k-1}\}\odot \{r+s,r^{k-1}\}\odot\ldots\odot\{r+s,r^{k-1}\}\ni\{d\},
\eqn
where $r\!=\!w$ and the inner product is taken $m$ times, subsequently written as $\odot^m\{r+s,r^{k-1}\}$.

For reasons discussed below, taking $r\!=\!0$ results in the trivial inner product:
\beqn
\{s\}\odot\{s\}=\{s\},\nonumber
\eqn
for all integers $s>0$. Extending to $m>2$,
\beqn
\odot^m\{s\}=\{s\},\nonumber
\eqn
and the corresponding Markov invariant is denoted as $\Phi$ with degree $d\!=\!1$ and weight $w\!=\!0$, and simply expresses the conservation of total probability under the action of the Markov semigroup:
\beqn
\Phi(P)\equiv \sum_{i_1,i_2,\ldots,i_m} P_{i_1i_2\ldots i_m}=1.\nonumber
\eqn
Here $\Phi$ is the invariant corresponding to $s\!=\!1$ and for $s\!>\!1$ the invariant is simply the power $\Phi^s$.

For fixed $m$, and any two invariants $f,f'$ of degree $d,d'$ and weight $w,w'$, we can form the pointwise product $f\cdot f'$ which is itself an invariant of degree $d\!+\!d'$ and weight $w\!+\!w'$. If $w\!=\!w'$, we can form an invariant from the sum $f\!+\!f'$. These statements establish that the invariants, ${\mathcal P}{(}V{)}{}^G$, form a \emph{graded ring} \cite{kelarev2002} (where the grading is over both the degree $d$ and the weights $w$). In particular, it is important to note that we can increase the degree of any invariant (keeping the weight fixed) by multiplying it with the trivial invariant $\Phi$.

When searching for Markov invariants, we must note that the sufficiency condition (\ref{eq:SuffCond}) will include these powers, and hence in what follows we must allow for this over-counting. In the conclusions we will expand upon this observation with some comments in regard to classifying the ring of invariants.

\subsubsection*{The general linear, or $s\!=\!0$, case}

Recalling Theorem 1, we see that for $s\!=\!0$, the Markov invariants are simultaneously invariants under the action of the general linear group. Taking $r\!=\!1$, the inner multiplication is trivial:
\beqn
\{1^k\}\odot\{1^k\}&=\{k\}.\nonumber
\eqn
This reflects that the Kronecker product of the \emph{alternating} representation of $\mathfrak{S}_k$, associated with the partition $(1^k)$, taken with itself, is the trivial representation, which in turn is associated with the partition $(k)$. Recall that the alternating representation is one-dimensional whose action on $\mathbb{C}$ defined as multiplication by $+1$ if $\sigma$ is an even permutation and is $-1$ otherwise. For this one-dimensional representation, the Kronecker product is simply the numeric product, with the result being the trivial representation where every permutation is mapped to $+1$. Similarly
\beqn
\{1^k\}\odot \{k\}=\{1^k\},\nonumber
\eqn
and we see that there exists a single Markov invariant of degree $d\!=\!k$ and weight $w\!=\!1$ for all \emph{even} values of $m$.  

A very familiar example occurs for $m\!=\!2$ where, as we will discuss in \S\ref{subsec:EvaluateTree}, the invariant arises as the Log-Det distance function \cite{steel1994}. In the next case, $m\!=\!4$, we refer to the corresponding Markov invariant as the \emph{quangle}.

Considering $m\!=\!2$ and $r\!=\!2$, we have
\beqn
\{2^k\}\odot\{2^k\}&\ni\{2k\},\nonumber
\eqn
for $2\leq k\leq 4$. For each $k$, these invariants can be accounted for by taking the previous invariant and multiplying by $\Phi$. Thus nothing new is gained.

However, taking $m\!=\!3$, it follows that there exists an invariant of degree $d\!=\!2k$ and weight $w\!=\!2$:
\beqn
\{2^k\}\odot\{2^k\}\odot\{2^k\}\ni\{2k\}.\nonumber
\eqn 
For $k\!=\!2$ this invariant is known in the quantum physics literature as the \emph{tangle} \cite{coffman2000,dur2000}, where it is drawn upon to classify entanglement in 3-qubit systems, and has been generalized for $k\!=\!3$ and 4 in the context of phylogenetics in \cite{sumner2006}. In \S\ref{subsec:EvaluateTree} we will briefly review the most striking properties of the tangle relevant to phylogenetics.

\subsubsection*{Bona-fide Markov invariants, $s\!>\!0$}

Here we consider the case $s\!>\!0$, where the resulting Markov invariants are \emph{not} simultaneously valid for the general linear group. In Table \ref{tab:inner} we present the number of weight $w\!=\!1$ invariants that exist for the cases $k\!=\!2,3,4$; $m\!=\!2,3,\ldots ,10$ and $s\!=\!1,2$. All required computations were performed using \texttt{Schur}, and we have not reduced for over-counting. 
\begin{table}[tbp]
\centering
\begin{tabular}[h]{|r||r|r|r|r|r|r|}
\hline
 & \multicolumn{2}{|c|}{$k\!=\!2$} & \multicolumn{2}{|c|}{$k\!=\!3$} & \multicolumn{2}{|c|}{$k\!=\!4$} \\
\hline
$m$ & $\{21\}$ & $\{31\}$ & $\{21^2\}$ & $\{31^2\}$ & $\{21^3\}$ & $\{31^3\}$ \\
\hline
2   & 			 1   &  			 1  &  			 1 &   			  1   &    				1 &       				1\\
3   & 			 1   &  			 1  &  			 1 &       		   1  &   				0 &       				1\\ 
4   &			     3   &  		 	4  &  			 4 &   			13   &    				4 &       				16\\
5   & 		 	 5   & 			10 & 				10 &  			 61  &   					 6 &      	 			137\\
6   & 			11  &  			31 & 				31 & 				397  &  					40 &       				1396\\
7   &      	    21  &  			91 & 				91 & 			2317    &				126   &					13881\\
8   &         	43  &			  274 &			  274  &		  14029    &				568   &					138916\\
9   &				85  &			  820 &			820    &       83917    & 			2142     &				1388857\\	
10 &			   171 &			2461 &			2461	 &      504013    &     		8824     &				13888996\\
\hline
\end{tabular}
\caption{Occurrences of $\{d\}$ in $\odot^m\{r+s,r^{k-1}\}$ with $rk\!+\!s\!=\!d$}
\label{tab:inner}
\end{table}
In Table \ref{tab:nomenclature} we summarize the Markov invariants for which we have successfully computed explicit polynomial forms. Here we also record the nomenclature we have developed. Presently we discuss the particular properties of these invariants when evaluated on phylogenetic tensors derived from a tree.
\subsection{What happens on a phylogenetic tree?}\label{subsec:EvaluateTree}

By definition, the expectation value of a (bias corrected) Markov invariant, $f$, depends only upon the internal part of the phylogenetic tree:
\beqn
E[\widehat{f}(Z)]=f(P)=(\det{M_1}\det{M_2}\ldots\det{M_m})^wf(\widetilde{P})\nonumber,
\eqn
where $Z$ is the observed counts of character patterns, $P$ is the phylogenetic tensor corresponding to the joint distribution on the tree, and the trimmed tensor $\widetilde{P}$, defined in \S\ref{sec:phylotens}, is formed by setting the lengths of the pendant edges to zero. It is exactly this property that can be exploited in the practical setting of reconstructing phylogenetic trees from molecular sequence data.

As discussed in the closing comments of \S\ref{subsec:RepGLK}, Markov invariants exist independently of any notion of a tree, and to uncover their potential use in the problem of phylogenetic tree reconstruction it becomes necessary to analyse their structure on particular trees. Crucial to this examination is the \emph{generalized pulley principle} presented in \cite{sumner2006}, which establishes that the family of probability distribution resulting from taking the general Markov model on a particular tree is unchanged under arbitrary placement of the root of the tree (see \cite{allman2003} for an equivalent discussion). Thus, our task is to search for algebraic relations between the Markov invariants valid for a given $m$, when evaluated upon the trimmed phylogenetic tensors corresponding to particular trees with $m$ leaves. We are free to place the root arbitrarily, and we choose to evaluate the Markov invariants on trees where the root is located to our convenience.
 
In Appendix~\ref{sec:YoungOps} we present the general procedure for computing the explicit polynomial form of Markov invariants using the Young's operators (\S\ref{subsec:RepGLK}) associated with the relevant partitions. Our general procedure was to take these explicit forms and then search for algebraic relations when the invariants are evaluated on the pruned tensor $\widetilde{P}$ defined by a particular tree. In the general case, any such relations potentially lead to phylogenetically informative statistics, valid under a general model of sequence evolution. 
Presently we will report upon this procedure in the known cases, $m=2,3$ and 4.
\begin{table}[tbp]
\centering
\begin{tabular}[h]{|c|c|l|l|c|}
\hline
Name & Symbol & Inner multiplication & Group & $(d,w)$ \\
\hline
det & $\text{Det}$ & $\odot^2\{1^2\}=\{2\}$ & $\times^2GL(2)$ & (2,1) \\
& & $\odot^2\{1^3\}=\{3\}$ & $\times^2GL(3)$ & (3,1) \\
& & $\odot^2\{1^4\}=\{4\}$ & $\times^2GL(4)$ & (4,1) \\
tangle & $T$ & $\odot^3\{2^2\}\ni\{4\}$ & $\times^3GL(2)$ & (4,2)\\
& & $\odot^3\{2^3\}\ni\{6\}$ & $\times^3GL(3)$ & (6,2) \\
& & $\odot^3\{2^4\}\ni\{8\}$ & $\times^3GL(4)$ & (8,2) \\
stangle & $T^{s}$  & $\odot^3\{21\}\ni\{3\}$ & $\times^3GL_1(2)$ & (3,1) \\
 && $\odot^3\{21^2\}\ni\{4\}$ & $\times^3GL_1(3)$ & (4,1) \\
&  & $\odot^3\{31^3\}\ni\{6\}$ & $\times^3GL_1(4)$ & (6,1) \\
quangle & $Q$ & $\odot^4\{1^2\}\ni\{2\}$ & $\times^4GL(2)$ & (2,1) \\
&  & $\odot^4\{1^3\}\ni\{3\}$ & $\times^4GL(3)$ & (3,1) \\
&  & $\odot^4\{1^4\}\ni\{4\}$ & $\times^4GL(4)$ & (4,1) \\
 squangle & $Q^s$  & $\odot^4\{21\}\ni 3\{3\}$ & $\times^4GL_1(2)$ & (3,1) \\
 &  & $\odot^4\{21^2\}\ni 4\{4\}$ & $\times^4GL_1(3)$ & (4,1) \\
&  & $\odot^4\{21^3\}\ni 4\{5\}$ & $\times^4GL_1(4)$ & (5,1) \\
\hline
\end{tabular}
\caption{Markov invariants of degree $d$ and weight $w$ for $m$ leaves, }
\label{tab:nomenclature}
\end{table}
\subsubsection*{The simplest Markov invariant: the Log-Det}

Recall that the generic phylogenetic tensor on $m\!=\!2$ leaves (Figure~\ref{fig:2leaf}) can be written in the form
\beqn
P=(M_1\otimes M_2)\cdot (\delta \cdot\pi).\nonumber
\eqn
The corresponding trimmed tensor, $\widetilde{P}\!=\!\delta\cdot \pi$, can be expressed in the $\delta_1,\delta_2,\ldots ,\delta_k$ basis with the components
\beqn
\widetilde{P}_{i_1i_2}=\delta_{i_1i_2}\pi_{i_1}.\nonumber
\eqn
As we showed above, there exists a single Markov invariant for $m\!=\!2$. The polynomial form of this invariant is easily derived by considering rank 2 tensors as matrices, and taking the determinant. Since the invariant is a function on tensors, we make the distinction by using a capital letter and denoting the invariant as $\text{Det}$. This distinction can be compared directly to the use of the determinant function in \cite{barry1987} as opposed to the use in \cite{steel1994}.
 
Substitution gives
\beqn
\text{Det}(\widetilde{P})=\prod_{1\leq i \leq k}\pi_i,\nonumber
\eqn
such that, by the definition of $\text{Det}$ as a Markov invariant,
\beqn\label{eq:DetProperty}
\text{Det}(P)=\det(M_1)\det(M_2)\prod_{1\leq i \leq k}\pi_i.
\eqn
This form holds for \emph{any} $k$, and is exploited by taking the logarithm and computing the Log-Det distance measure \cite{lake1994,lockhart1994}.
\begin{figure}[tbp]
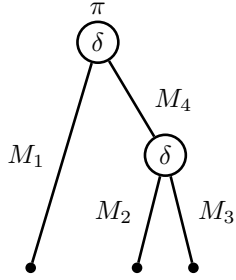

    \centering
    \rput[br](1,0.75){$\pi$}
    \pstree{\Tcircle{$\delta$}}{%
        \skiplevel{\Tdot \tlput{$M_{1}$}}
        \Tn
        \pstree{\Tcircle{$\delta$} \trput{$M_{4}$}}{
            \Tdot \tlput{$M_{2}$}
            \Tdot \trput{$M_{3}$}
        }
    }
    \caption{Phylogenetic tensor for the tree (1,23)}
    \label{fig:3leaf}
\end{figure}

\subsubsection*{Triplet distances: the tangle}

Inspection of Table~\ref{tab:inner} reveals that for $m\!=\!3$ and $s\!=\!0$ there exists a Markov invariant, for each of $k\!=\!2,$ 3 and 4, of degree $d\!=\!2k$ and weight $w\!=\!2$.
This invariant is valid for phylogenetic trees with three leaves.
For each of $k\!=\!2,$ 3 and 4, the explicit polynomial forms of the tangle are basis independent (by definition) and have 12, 1152 and 431424 terms respectively.

The generic phylogenetic tensor on the three leaf tree (Figure~\ref{fig:3leaf}) can be expressed as
\beqn\label{eq:3leaf}
P=(1\otimes M_2\otimes M_3)\cdot (1\otimes \delta)\cdot (M_1\otimes M_4) \cdot (\delta \cdot\pi).
\eqn
The trimmed tensor, $\widetilde{P}=(1\otimes \delta)\cdot (1\otimes M_4) \cdot (\delta \cdot\pi)$, has components
\beqn\label{eq:clipped3}
\widetilde{P}_{i_1i_2i_3}=P^\ast_{i_1i_2}\delta_{i_2i_3},
\eqn
where $P^\ast\!=\!(1\otimes M_4) \cdot (\delta\cdot \pi)$ is the pruned tensor.

The tangle is a Markov invariant and hence satisfies
\beqn
T(P)=(\det M_1\det M_2\det M_3)^2T(\widetilde{P}).\nonumber
\eqn
By explicit computation we have found that, for each of $k\!=\!2,$ 3 and 4,
\beqn
T(\widetilde{P})=\text{Det}^2(P^\ast).\nonumber
\eqn
Thus we see that the induced function of the tangle is $T^\ast\equiv\text{Det}^2$. This is the example we promised in \S\ref{subsec:MarkovInvariantsDef}.

Consistent with (\ref{eq:DetProperty}) we have
\beqn
\text{Det}(P^\ast)=\det M_4\left(\prod_{1\leq i \leq k}\pi_i\right),\nonumber
\eqn
so, finally, we see that
\beqn\label{eq:tangleProperty}
T(P)=(\det M_1\det M_2\det M_3\det M_4)^2\left(\prod_{1\leq i \leq k}\pi_i\right)^2.
\eqn
Due to the generalized pulley principle, (\ref{eq:tangleProperty}) holds for the phylogenetic tensor corresponding to \emph{any} tree with three leaves. Comparing directly to (\ref{eq:DetProperty}) it is clear that the tangle may be used similarly to the Log-Det pairwise distance but for triplets of molecular sequence data.  For further details in this direction see \cite{sumner2006}.

\subsubsection*{Informative statistic: the stangle}

We see from Table~\ref{tab:inner} that for $m\!=\!3$ and $s\!=\!2$ there also exists, for each $k=2$, 3 and 4, a weight $w\!=\!1$ Markov invariant valid for trees with three leaves (of degree $d\!=\!6$ for $k\!=\!4$ states). 
We refer to this invariant as the \emph{stangle}, that is, the \emph{s}tochastic \emph{tangle} (see \cite{sumner2006a} for explicit expressions for the $k\!=\!2$ and 3 cases).
As discussed in Appendix~\ref{sec:YoungOps}, the explicit polynomial form of the stangle for $k\!=\!4$ is known only in a basis different from the standard $\delta_1,\delta_2,\ldots,\delta_k$.
In this basis, the stangle has 1404 terms with relevant data files available on Charleston's website \cite{squangle}.
This does not, however, prevent us from using the stangle in a practical setting as evaluation can be performed in this basis by transforming the \emph{data set} (pattern counts) into the required basis.

For the trimmed tensor with components given by (\ref{eq:clipped3}), explicit computation shows that the stangle satisfies $T^{s}(\widetilde{P})\equiv 0$.
Thus the stangle is simultaneously a \emph{phylogenetic invariant} for a tree with three leaves (of course this again holds for \emph{any} tree with three leaves). 

Given an unbiased estimator, $\widehat{T}^s$, of the stangle, we see that under the family of probability distributions described by (\ref{eq:3leaf}), the expectation value of this estimator  when evaluated on triplets of aligned DNA sequences is zero :
\beqn
E[\widehat{T}^s(Z)]=0,\nonumber
\eqn
where $Z$ is the tensor of observed pattern counts in the aligned sequence data. Deviation from zero by the observed value of the stangle can thus be viewed as evidence that the data set violates the assumptions of the Markov model. We have had some preliminary (unpublished) success capitalizing on this property to rank subsets of aligned molecular sequences according to apparent concurrence with model assumptions.

Note that the stangle must occur within the framework of phylogenetic invariants presented in \cite{allman2003} and the discussion of \cite{landsberg2006}. 
It would be interesting to determine whether the stangle is a linear combination (with coefficients that are $d\!=\!1$ polynomials) of the quintic phylogenetic invariants presented in \cite{sturmfels2007}. 
However, whether or not this is the case is beyond the theoretical techniques presented in this paper and more work needs to be done before the precise connections between the stangle and the known phylogenetic invariants for this case become transparent.
Further, because the explicit polynomial form of the stangle in the standard basis is not known, brute-force determination is impractical using algorithms presently known to the authors. 

\subsubsection*{Quartet inference: the squangles}

Inspection of Table~\ref{tab:inner} reveals that for $k\!=\!4$ and $m\!=\!4$, there exist four Markov invariants of degree $d\!=\!5$ and weight $w\!=\!1$ relevant to phylogenetic trees with four leaves. We refer to these invariants as the \emph{squangles}. 
Again, the explicit polynomial form of the squangles is known only in a basis different from the standard one, and data files can be found on Charleston's website \cite{squangle}.
We have found that three particular linear combinations of the squangles are tree informative. 
Here we denote these three squangles as $Q_1,Q_2$ and $Q_3$.
In the non-standard basis, $Q_1$ has 77004 terms, whereas both $Q_2$ and $Q_3$ have 91620 terms. 

\begin{figure}[tbp]
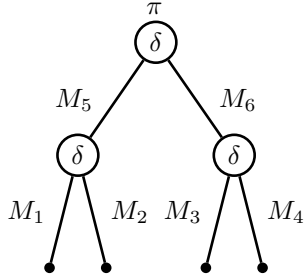

    \centering
    \rput[br](1.5,0.75){$\pi$}
    \pstree{\Tcircle{$\delta$}}{%
        \pstree{\Tcircle{$\delta$} \tlput{$M_{5}$}}{
            \Tdot \tlput{$M_{1}$}
            \Tdot \trput{$M_{2}$}
        }
        \Tn
        \pstree{\Tcircle{$\delta$} \trput{$M_{6}$}}{
            \Tdot \tlput{$M_{3}$}
            \Tdot \trput{$M_{4}$}
        }
    }
    \caption{Phylogenetic tensor for the tree (12,34)}
    \label{fig:4leafA}
\end{figure}

On the quartet tree in Figure~\ref{fig:4leafA},  the generic phylogenetic tensor is
\beqn
P=(M_1\otimes M_2\otimes M_3\otimes M_4)\cdot (\delta\otimes \delta)\cdot (M_5\otimes M_6)\cdot (\delta\cdot \pi).\nonumber
\eqn
The trimmed tensor $\widetilde{P}=(\delta\otimes \delta)\cdot (M_5\otimes M_6)\cdot (\delta\cdot \pi)$ has components:
\beqn
\widetilde{P}_{i_1i_2i_3i_4}=P^\ast_{i_1i_3}\delta_{i_1i_2}\delta_{i_3i_4},\nonumber
\eqn
with the pruned tensor given by $P^\ast=(M_5\otimes M_6)\cdot  (\delta\cdot \pi)$. This form of the trimmed tensor can be evaluated directly on the explicit polynomial form of the squangles. We found that on the tree $(12,34)$ the squangles satisfy the algebraic relations:
\beqn
Q_1(\widetilde{P})=0,\quad Q_2(\widetilde{P})=-Q_3(\widetilde{P})>0,\nonumber
\eqn
with, intriguingly, the polynomial form of $Q_2(\widetilde{P})$ with respect to the components $P^\ast_{i_1i_2}$ taking that of the \emph{permanent}\footnote{The permanent has identical algebraic form to the determinant of a matrix but with each term replaced by its absolute value.}, which, unfortunately for the phylogenetic context, is not a Markov invariant. 
 
\begin{figure}[tbp]
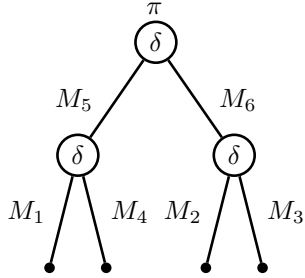

    \centering
    \rput[br](1.5,0.75){$\pi$}
    \pstree{\Tcircle{$\delta$}}{%
        \pstree{\Tcircle{$\delta$} \tlput{$M_{5}$}}{
            \Tdot \tlput{$M_{1}$}
            \Tdot \trput{$M_{4}$}
        }
        \Tn
        \pstree{\Tcircle{$\delta$} \trput{$M_{6}$}}{
            \Tdot \tlput{$M_{2}$}
            \Tdot \trput{$M_{3}$}
        }
    }
    \caption{Phylogenetic tensor for the tree (14,23)}
    \label{fig:4leafC}
\end{figure}

An identical procedure was carried out on the phylogenetic tensors corresponding to the trees in Figure~\ref{fig:4leafB} and Figure~\ref{fig:4leafC}. This produced the relations
\begin{alignat}{2}
Q_2(\widetilde{P})&=0,&\quad \quad Q_1(\widetilde{P})&=Q_3(\widetilde{P})>0,\nonumber\\
Q_3(\widetilde{P})&=0,&\quad -Q_1(\widetilde{P})&=-Q_2(\widetilde{P})>0,\nonumber
\end{alignat}
respectively. 

\begin{figure}[tbp]
    \centering
    \rput[br](1.5,0.75){$\pi$}
    \pstree{\Tcircle{$\delta$}}{%
        \pstree{\Tcircle{$\delta$} \tlput{$M_{5}$}}{
            \Tdot \tlput{$M_{1}$}
            \Tdot \trput{$M_{3}$}
        }
        \Tn
        \pstree{\Tcircle{$\delta$} \trput{$M_{6}$}}{
            \Tdot \tlput{$M_{2}$}
            \Tdot \trput{$M_{4}$}
        }
    }
    \caption{Phylogenetic tensor for the tree (13,24)}
    \label{fig:4leafB}
\end{figure}

Noting these relations, we see that we have constructed tree-informative phylogenetic invariants for trees with four leaves. In particular, for the unbiased estimators thereof, we have
\beqn
E[\widehat{Q}_1(Z)]\equiv 0,\qquad E[\widehat{Q}_2(Z)+\widehat{Q}_3(Z)]\equiv 0;\nonumber
\eqn
for the tree $(12,34)$,
\beqn
E[\widehat{Q}_2(Z)]\equiv 0,\qquad E[\widehat{Q}_1(Z)-\widehat{Q}_3(Z)]\equiv 0;\nonumber
\eqn
for the tree $(13,24)$, and
\beqn
E[\widehat{Q}_3(Z)]\equiv 0,\qquad E[\widehat{Q}_1(Z)-\widehat{Q}_2(Z)]\equiv 0;\nonumber
\eqn
for the tree $(14,34)$. 
We also note that the linear combination
\beqn
W:=Q_1-Q_2-Q_3,\nonumber
\eqn
satisfies $E[\widehat{W}(Z)]\equiv 0,$ for \emph{any} phylogenetic tree with four leaves.

\begin{figure}[btp]
\centering
\scalebox{.45}{\includegraphics{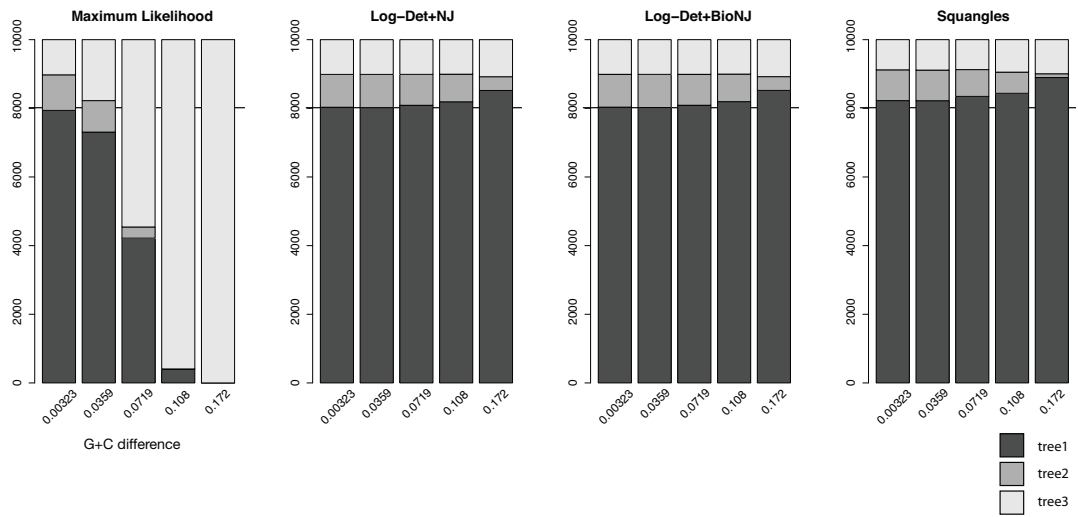}}
\caption{\textbf{Quartet reconstruction using the squangles.}  The charts present how many times the \texttt{tree1}=(12,34), \texttt{tree2}=(13,24) and \texttt{tree3}=(14,23) were reconstructed using each of the three methods displayed. The tree used to simulate the data was \texttt{tree1}.}
\label{fig:dunce}
\end{figure}

The bar charts in Figure~\ref{fig:dunce} compare the success of three tree inference methods tested on data sets created using \texttt{Hetero} \cite{hetero}. 
All parameter settings used are as presented in \cite{jermiin2004} with sequence length $N$=10000 and 10000 runs being completed in each case. 
A molecular clock was imposed, and for each run the tree used to simulate the data was
\beqn
\texttt{tree1}=((Seq1:0.495,Seq2:0.495):0.005,(Seq3:0.495,Seq4:0.495):0.005),\nonumber\\ 
\eqn
with branch lengths given in time units and $t\!=\!0.495$ and .005 corresponding to 0.1485 and 0.0015 expected number of state changes, respectively.
The G+C content was made to increase in leaves 1 and 4 and was reduced in leaves 2 and 3. 
This tends to bias tree inference to \texttt{tree3}$=(14,23)$ as sequences 1 and 4 will tend to be more similar purely because of the G+C content.

The Maximum Likelihood and Log-Det+NJ quartet inferences were performed using the default settings in \texttt{Phylip} \cite{phylip}, whereas the Log-Det+BIONJ inferences were performed using the \texttt{R} \cite{Rproject} package ``\texttt{ape}'' \cite{ape}.
Finally, the squangles inferences  were implemented in \texttt{R} using our own original code \cite{squangle}.
For the purpose of making a rough comparison, on average each evaluation took .58s for maximum likelihood, .036s for Log-Det+NJ, .090s for Log-Det+BIONJ, and .085s for the squangles.
The squangles routine was designed for illustrative purposes only and was performed under simple statistical assumptions, as follows.

The squangles were taken to be stochastically independent and normally distributed, with identical variances, $\sigma^2$, and mean values set to 0 or $u>0$, depending on the quartet under consideration and the expectation values given above. 
That is, for each quartet in turn, we took
\begin{itemize}
\item[] $\mathbb{P}[Q_1,Q_2,Q_3|(12,34)]\sim \mathcal{N}(0,\sigma^2)*\mathcal{N}(u,\sigma^2)*\mathcal{N}(-u,\sigma^2)$,
\item[] $\mathbb{P}[Q_1,Q_2,Q_3|(13,24)]\sim \mathcal{N}(u,\sigma^2)*\mathcal{N}(0,\sigma^2)*\mathcal{N}(u,\sigma^2)$,
\item[] $\mathbb{P}[Q_1,Q_2,Q_3|(14,23)]\sim \mathcal{N}(-u,\sigma^2)*\mathcal{N}(-u,\sigma^2)*\mathcal{N}(0,\sigma^2)$.
\end{itemize}
Our primary scientific justification for these assumptions is that the resulting quartet inference routine performs rather well.

Under these assumptions the maximum likelihood estimate (MLE) of $u$ is independent of $\sigma^2$, and is equivalent to the least squares estimator.
\emph{Analytic} solutions are easily derived:
\beqn
\text{MLE}\left[u|(12,34)\right]&=\max\left[0,\frac{Q_2(Z)-Q_3(Z)}{2}\right],\nonumber\\
\text{MLE}\left[u|(13,24)\right]&=\max\left[0,\frac{Q_1(Z)+Q_3(Z)}{2}\right],\\
\text{MLE}\left[u|(14,23)\right]&=\max\left[0,\frac{-(Q_1(Z)+Q_2(Z))}{2}\right].
\eqn
For each data set and candidate quartet, we computed the MLE for the mean value $u$ and chose the quartet with the maximum likelihood. 

While our demonstration is not intended as an exhaustive comparison between the performance of our method and ML using the default settings of \texttt{Phylip}, it does show that using a stationary model for ML can lead to incorrect tree inference if the data was produced by a non-stationary process.
With that caveat, it is clear that ML performs very badly as the G+C content increases, strongly favouring \texttt{tree3}. 
The Log-Det routine is robust against varying G+C content as the technique is based on a general model (this is consistent with what was found in \cite{jermiin2004}). 
Being valid for a general model, the squangles are also robust against varying G+C content, and actually perform slightly better than Log-Det.

Interestingly, as the G+C content increases, the Log-Det and the squangles infer the true tree \textit{more} often. 
Careful inspection reveals that this is because, as the G+C content increases, both these techniques tend to infer \texttt{tree2} less often and \texttt{tree3} at approximately the same rate, favouring \texttt{tree1}. 
This effect is more pronounced for the squangles.

\subsection{Mixed weight Markov invariants}\label{subsec:MixedWeight}
Here we report upon the existence of some mixed weight invariants for various cases of interest to phylogenetics. 
The polynomial form and algebraic structure on trees of these invariants remains completely unexplored.

We concentrate on $k\!=\!4$ and look for mixed weight invariants for the degree $d\!=\!8$  partition shapes $\{2^4\}$ and $\{51^3\}$, corresponding to $s\!=\!0$ and 4 respectively.

In the $m\!=\!2$ case, we find that
\beqn
\{2^4\}\odot\{51^3\}\nonumber
\eqn
does not contain $\{8\}$, which means there does not exist a mixed weight invariant for trees on two leaves.

In the $m\!=\!3$ case, we have
\beqn
\{2^4\}\odot\{51^3\}\odot\{51^3\}\ni \{8\},\nonumber\\
\{2^4\}\odot\{2^4\}\odot\{51^3\}\ni \{8\}.
\eqn
Writing $w=(w_1,w_2,w_3)$, we see that, including the three possible permutations across the inner products,  there exist mixed weight invariants for the cases $w\!=\!(2,1,1),(1,2,1),(1,1,2)$ and $w\!=\!(2,2,1),(2,1,2),(1,2,2)$ respectively.

In the $m\!=\!4$ case, we have
\beqn
\{2^4\}\odot\{51^3\}\odot\{51^3\}\odot\{51^3\}&\ni 14\{8\},\nonumber\\
\{2^4\}\odot\{2^4\}\odot\{51^3\}\odot\{51^3\}&\ni 9\{8\},\\
\{2^4\}\odot\{2^4\}\odot\{2^4\}\odot\{51^3\}&\ni 4\{8\}.
\eqn
Taking account of the permutations, we see that there exist $14\!\times\! 4=54$, $9\!\times\! 6=54$ and $4\!\times\!4=16$ mixed weight invariants for the cases $w\!=\!(2,1,1,1)$, $w\!=\!(2,2,1,1)$ and $w\!=\!(2,2,2,1)$ respectively.

Finally, in the $m\!=\!5$ case, we have
\beqn
\{2^4\}\odot\{51^3\}\odot\{51^3\}\odot\{51^3\}\odot\{51^3\}&\ni 527\{8\},\nonumber\\
\{2^4\}\odot\{2^4\}\odot\{51^3\}\odot\{51^3\}\odot\{51^3\}&\ni 212\{8\},\\
\{2^4\}\odot\{2^4\}\odot\{2^4\}\odot\{51^3\}\odot\{51^3\}&\ni 90\{8\},\\
\{2^4\}\odot\{2^4\}\odot\{2^4\}\odot\{2^4\}\odot\{51^3\}&\ni 46\{8\}.
\eqn
Again taking account of the permutation, we see that there exist $527\!\times\! 5=2635$, $212\!\times\! 10=2120$, $90\!\times\! 10=900$ and $46\!\times\! 5=230$ mixed weight invariants for the cases $w\!=\!(2,1,1,1,1)$, $w\!=\!(2,2,1,1,1)$, $w\!=\!(2,2,2,1,1)$ and $w\!=\!(2,2,2,2,1)$ respectively.

We expect that a future analysis of the explicit form of these invariants will lead to quite an array of informative statistics for phylogenetics. 

%% file: conc.tex
\section{Discussion}
\label{sec:Outlook}
In this work we have defined and explored the construction of `Markov invariants'. The primary tool exercised was group representation theory, applied to the usual Markov process present in probabilistic models of phylogenetic trees.

It is evident that our present approach to phylogenetics offers many possibilities for further study. 
The various Markov invariants that we have identified and constructed provide strong candidates for improved tree estimation and parameter recovery under general model assumptions. 
In particular, the stangle (\S\ref{subsec:EvaluateTree}) seems to provide a robust indicator of phylogenetic signal in subsets of aligned molecular sequences.
Efforts are underway to incorporate the stangle into a clustering algorithm that provides the means to divide large phylogenetic data sets into smaller, manageable parts, inspired, in part, by the Disk-Covering technique of \cite{huson1999}. 
In \S\ref{subsec:EvaluateTree}, we presented a Markov invariant based quartet inference technique. 
The maximum likelihood estimation we employed was based on rather simple statistical assumptions, and it is clear that this technique could easily be improved upon.
Detailed knowledge of the invariants' distribution is desirable not only in order to achieve correct tree inferences, but also to find confidence intervals (as \cite{massingham2007} do for Log-Det and ML distances).
In all its glorious detail, the joint distribution of the squangles can be derived using the multinomial distribution of $Z$:
\beqn
\mathbb{P}[Q_1(Z)\!=\!q_1,Q_2(Z)\!=\!q_2,Q_3(Z)\!=\!q_3]&=\sum_{z\in \Upsilon}\mathbb{P}[Z\!=\!z;N]=N!\sum_{z\in \Upsilon}\left(\prod_{I\in K^m}\frac{\mu_I^{z_{I}}}{z_I!}\right),\nonumber
\eqn 
where the summation is over the variety $\Upsilon\!:=\!\{z|Q_1(z)-q_1\!=\!Q_2(z)-q_2\!=\!Q_3(z)-q_3\!=\!0\}$.
However, this distribution depends implicitly upon the model parameters underlying $\mu$, and therefore negates the whole point of employing invariants in the first place!
Clearly, a more coarse grained approach is desirable for intuiting an approximate distribution for the invariants that depends on just a few shape parameters.
This can be achieved variously by studying the relevance and impact of the central limit theorem, deriving the first few moments using the generating function (\ref{generatingfunction}), or conducting extensive simulation studies.
This would help to provide rigorous justification for taking the invariants as normally distributed, as we did for the squangles in \S\ref{subsec:EvaluateTree}. 

Citing poor performance on short sequences \cite{hillis1994,huelsenbeck1995}, there is a somewhat popular opinion that phylogenetic invariants are of limited utility when it comes to phylogenetic inference in practice.
However, recent work suggests that this performance can be greatly improved by identifying ``powerful'' invariants \cite{eriksson2008}.
For instance, \cite{casanellas2007} chose invariants for the K3ST model using a criterion arising from algebraic geometry, and \cite{eriksson2008b} used a learning algorithm to choose invariants for the K3ST and Jukes-Cantor models.
However, determining criteria that guarantee identification of statistically powerful invariants is in general an outstanding problem. 
In this context, we have shown clearly that Markov invariants can be of significant practical utility.
For instance, one need only note that the simplest Markov invariant forms the structure of the Log-Det distance measure, an extremely popular tool employed in countless phylogenetic studies, while the simulation study we presented in \S\ref{subsec:EvaluateTree} shows that Markov invariants can be used to infer quartet phylogenies with a success rate equivalent to, or greater than, popular methods.
For phylogenetic invariants that arise as Markov invariants, it would be interesting to determine whether the additional analytic structure imposed by group invariance provides an effective criterion for identification of powerful invariants.


Markov invariants occur as one-dimensional representations of a group action associated with the Markov semigroup. 
In this regard, we applied only a particular instance of the group branching rule (\ref{eq:LeafBranchingRule}) that requires each of the irreducible modules to be one-dimensional. 
The standard approach to maximum likelihood exploits the trivial instance of the same branching rule with $\{\lambda\}\!=\!\{\sigma_1\}\!=\! \{\sigma_2\}\!=\!\ldots\!=\! \{\sigma_m\}\!\equiv \!\{1\}$, taking $m$ copies of the $k$-dimensional \emph{defining} representation to obtain a $k^m$-dimensional and degree $d\!=\!1$ polynomial representation.
From this perspective, the standard approach and the Markov invariants are simply two cases where the transformation properties of polynomials of molecular sequence data under the time evolution (\ref{eq:PhylogeneticTreeModel}) are exploited. 
This begs the question whether there exist polynomial representations, of dimension other than these two extremes, that can also be effectively utilized in practical phylogenetic tree inference.

Many of the different classes of phylogenetic models \cite{huelsenbeck2004} can be affiliated with appropriate subgroups of $GL(k)$, and can therefore be expected to have a place in the subgroup chain (\ref{eq:subgroupchains}). 
In principle we can modify Theorem~1 (\S\ref{subsec:theorem}) for each of these models and construct their associated Markov-type invariants. 
In this vein, Appendix~\ref{sec:K3ST} outlines a group-theoretic analysis of the Kimura 3ST model connecting the Hadamard conjugation with the construction of the Cartan subalgebra for this model.
The same considerations apply in principle to amino acid sequence models: this is simply a matter of setting $k\!=\!20$ and using the same theory, though the computations involved will of course be more lengthy.

As noted in \S\ref{subsec:RepGLK}, a representation-theoretic analysis of the space of phylogenetic tensors that includes the underlying tree structure has not been developed in this work. 
Ideally, for a given tree, one would like to obtain the structure of the ring of Markov invariants as a theoretical outcome, rather than obtain this structure using the post-hoc procedure presented in \S\ref{subsec:EvaluateTree}. 
A possible direction in this regard is to consider, for each tree with $m$ labelled leaves, the subgroup of $\mathfrak{S}_m$ induced by identifying permutations that leave the leaf labelling invariant. 
This subgroup is discussed in \cite[Chap. 2]{semple2003} and, in a different context, in \cite[Chap. 12, Topic 3]{biedenharn1981}. 
We conjecture that this group may play a role, analogous to that of the symmetric group for the Schur-Weyl duality, in the construction of the irreducible modules for the space of phylogenetic tensors.

The phylogenetic invariants form an \emph{ideal} in the associated polynomial ring, and hence Hilbert's basis theorem for finite-generatedness applies.
However, whether Markov invariants are finitely-generated is unknown. 
Technically, the group $GL_1(k)$ is \emph{non-reductive} (its finite-dimensional representations are not completely reducible). 
In the non-reductive case, standard theorems, such as finite-generatedness of polynomial invariant rings, do not apply. 
Thus it is unlikely that the Markov-type invariants will be finitely generated in general. 
A notable exception is provided by Weitzenb\"{o}ck's theorem \cite{seshadri1962,weitzenbock1931} for finite-dimensional representations of one-dimensional Lie groups. 
Continuing the subgroup chain (\ref{eq:subgroupchains}) to its natural limit, it follows that in the case of phylogenetic tensors, the group $GL_1(k)$ provides, on restriction, an indecomposable representation of the additive group $\mathbb{R}^+$ (corresponding to time evolution).
Thus, Weitzenb\"{o}ck's theorem is relevant to the analysis of Markov invariants in the current context.
In fact, this observation is pertinent to phylogenetic invariants for continuous time models, as they would occur as \emph{syzygies} \cite[Chap. 2]{olver2003} between invariants belonging to the generating set of the invariant ring for the representation of $\mathbb{R}^+$ in question.

\medskip
\noindent
\textbf{Acknowledgments} 

\noindent
This paper is the culmination of several years work resulting from interactions with many researchers in what is, for Sumner and Jarvis, an unfamiliar field. There are many people to acknowledge for their various contributions ranging from simple positive encouragement to important technical insights. In this regard, we would like to thank Michael Baake, Peter Forrester, Alexei Drummond, David Bryant, Mike Steel, David Penny, Mike Hendy, Susan Holmes, Mark Pagel, Andreas Dress, Elizabeth Allman, John Rhodes, John Robinson, Bertfried Fauser, Ron King, Mike Eastwood, Jim Bashford, Malgorzata O'Reilly, and Simon Wotherspoon. 
\newline

\noindent
This research was conducted with support from the Australian Research Council grants DP0344996, DP0770991 and DP0877447.

%% file: append.tex
\begin{appendix}
\setcounter{equation}{0}
\renewcommand{\theequation}{{A}-\arabic{equation}}

\section{Proof of Theorem 1}
\label{sec:Proofs}

We provide a tensor-based completion of the proof of Theorem 1 (\S\ref{subsec:theorem}), regarding the identification of one-dimensional irreducible representations of the groups $GL(k)$, $GL_1(k)$ and $GL_{1,1}(k)$.

Following the notation of \S\ref{sec:sec1},  a probability measure can be written in a basis of point measures, $\mu = \sum_{1\leq i\leq k} \mu_i \delta_i$, with the Markov semigroup acting as
\begin{align}\label{eq:MSGtrans}
\mu \mapsto   M\mu, \qquad
M\delta_i = & \, \sum_{1\leq j\leq k} \delta_j M_{ji}, \qquad
M\mu = \sum_{1\leq i\leq k}  {\mu '}_i \delta_i, \qquad 
{\mu '}_i = \sum_{1\leq j\leq k} M_{ij} \mu_j .
\end{align}
Moreover, probability conservation requires the column-sum condition $\sum_{1\leq i\leq k} M_{ij}\!=\!1,$ for all $1\leq j \leq k$. As discussed in \S\ref{sec:Groups}, this  affiliates the linear transformations $M\in\mathfrak{M}(k)$ with the subgroup $GL_1(k)\lhd GL(k)$. Correspondingly, a higher rank tensor, $\psi$, transforms under the action of $g\in GL_1(k)$, $\psi \mapsto {\psi'}$, with
\begin{align}
{\psi\,{}'}_{i_1i_2i_3 \ldots} =\sum_{1\leq j_1,j_2,j_3\ldots\leq  k}
g_{i_1j_1}g_{i_2j_2}g_{i_3j_3} \ldots \psi_{j_1j_2j_3\ldots }\, .
\nonumber
\end{align}
In order to find combinations of $\psi_{i_1i_2i_3i_4 \ldots}$ which remain invariant up to scaling under the $GL_1(k)$ action, we transform to a more convenient basis in which the distinguished role of the vector 
$(1,e^\top) = (1,1,\ldots 1)$ is identified. Following \cite{mourad2004}, define a nonsingular $k \times k $ matrix, $X$, with $1 \times 1 + (k\!-\!1)\times (k\!-\!1)$ block decomposition:
\begin{align}
X := & \, \left( \begin{array} {cc} 1 & e^\top \\ \eta & x \end{array} \right).
\end{align}

\medskip\noindent
\textbf{Lemma:} With respect to the similarity transformation $g \mapsto \widetilde{g} = X g X^{-1}$ defined by any fixed $X$ of the above form, $GL_1(k)$ is isomorphic to the affine group $A(k)\cong GL(k\!-\!1) \ltimes T(k\!-\!1)$. Furthermore, under the same mapping subject to the constraint $\eta = - x \cdot e$, $GL_{1,1}(k)$ is isomorphic to the group $GL(k\!-\!1)$.

\medskip\noindent
\textbf{Proof:}
Check explicitly that
\begin{align}
\mbox{if} \qquad g =& \,  \left( \begin{array} {cc} \lambda & \ell_1^\top \\ \ell_2 & m \end{array} \right),
\qquad \mbox{then} \qquad X g X^{-1} =   \left( \begin{array} {cc} 1 & 0 \\ \widetilde{\ell} & \widetilde{m} \end{array} \right), \nonumber
\end{align}
using the column-sum condition on $g$. Clearly, $\det{g} = \det{\widetilde{m}}$, so $\widetilde{m}\in GL(k\!-\!1)$ for all such $X$. Finally, if $\ell_2 = 1 - m \cdot e$, $\lambda = 1 - \ell_1^\top \cdot e$ and $\eta = - x\cdot e$, then $\widetilde{\ell} = 0$ in $X g X^{-1}$ and $g \in GL_{1,1}(k)$ is thereby identified with the $GL(k\!-\!1)$ subgroup of $GL(k)$ consisting of matrices in block form as displayed. \hfill  $\square$

\medskip\noindent
It is convenient to re-label the basis as $X\delta_{1} \!:=\! \widetilde{\delta}_0\!=\!\delta_1+\delta_2+\ldots+\delta_k$, $X \delta_a := \widetilde{\delta}_a$,
$a = 2, 3, \ldots, k$. In the new basis, probability measures will transform \emph{in}homogeneously, with the $\widetilde{\delta_0}$ components invariant; for example mimicking (\ref{eq:MSGtrans})
\begin{align}\label{eq:tildeTrans}
\widetilde{\mu}\,{}'_0 = & \,  \widetilde{\mu}_0,
\qquad \widetilde{\mu}\,{}'_a =  \widetilde{\ell}_a \,\widetilde{\mu}_0 + \sum_{b=2}^{k} \widetilde{m}_{ab} \, \widetilde{\mu}_b \, , 
\end{align}
and in this way we can deduce the transformation properties of higher-rank tensors.

As we discussed in \S\ref{subsec:RepGLK}, the finite dimensional \emph{irreducible} representations of $GL(k)$ associated with partitions $\lambda$, are realized by tensors of rank $|\lambda|$ whose indices satisfy particular symmetrization conditions to be outlined in Appendix~\ref{sec:YoungOps}: symmetrize across the rows and then anti-symmetrize down the columns of the associated standard tableau $\mathfrak{T}$. Conventionally, for example, we write for such a tensor the components
$ \psi_{{[}i_1 i_2 \ldots {]} {[}j_1 j_2 \ldots {]}{[} \ldots {]}} $.
Here the indices enclosed in braces ${[} \ldots {]}$ are mutually anti-symmetric, corresponding to column entries in $\mathfrak{T}$, and there are further cyclic identities (we need not consider) reflecting row dependencies of $\psi$. 

Below we will discuss properties of such tensors in the $\widetilde{\delta_0},\widetilde{\delta}_2,\ldots \widetilde{\delta}_{k}$ basis under the transformation (\ref{eq:tildeTrans}).
The crucial result will depend absolutely on the indices, and the symbol `$\psi$' will be superfluous. Hence, for ease of reading we will suppress the `$\psi$':
\beqn
 \psi_{{[}i_1 i_2 \ldots {]} {[}j_1 j_2 \ldots {]}{[} \ldots {]}}\equiv [}i_1 i_2 \ldots {]} {[}j_1 j_2 \ldots {]}{[} \ldots {].\nonumber
\eqn
This is consistent with the amusing comments in the preface of \cite{mccullagh1987}.

Consider the reduction of an irreducible representation $\lambda$ of $GL(k)$ with respect to the subgroup $GL(k\!-\!1)$ (equivalent, by the Lemma above, to considering the restriction to $GL_{1,1}(k)$ affiliated to the doubly-stochastic Markov semigroup). The partition labels $\overline{\lambda}$ of irreducible representations of $GL(k\!-\!1)$ arising from this restriction are related to those of $\lambda$ by the standard betweenness conditions 
\cite[Chap. V, \S 18]{weyl1950} (see also \cite{biedenharn1990,whippman1965}):
\begin{align}
\label{eq:Betweenness}
\lambda_1 \ge \overline{\lambda}_1 \ge & \ldots \ge \overline{\lambda}_{n-1}\ge \lambda_n.
\end{align}
Our present purpose is to identify one-dimensional representations of $GL(k\!-\!1)$, that may extend to one-dimensional representations of $GL_1(k) \cong GL(k\!-\!1)\ltimes T(k\!-\!1)$. Such tensor representations must be associated with partitions $\overline{\lambda} = (r^{k\!-\!1})$ all of whose columns have length $k\!-\!1$ corresponding to the $r^\text{th}$ power of the representation $M \mapsto \det{M}$. However, for such a $\overline{\lambda}$,
(\ref{eq:Betweenness}) above immediately implies that
\begin{align}
\lambda = & \, (r + s, r^{k-2}, t), \qquad \mbox{for some} \quad s \ge 0, \, t \le r, \nonumber
\end{align}
and we have established part (iii) of Theorem~1.

Within such tensor representations of type $(r + s, r^{k-2}, t)$, the component associated with the scalar representation $(r^{k\!-\!1})$ of $GL(k\!-\!1)$ is clearly
\begin{align}
{{[}0 a_{12} \ldots a_{1k}{]} {[}0 a_{22} \ldots a_{2k}{]} \ldots
[0 a_{t2} \ldots a_{tk}]
{[} b_{11} b_{12} \ldots b_{1,k\!-\!1}{]} \ldots {[} b_{r-t,1} b_{r2} \ldots b_{r-t,k\!-\!1}{]}0_1 0_2 \ldots 0_s}.
\nonumber
\end{align}
However, under the inhomogeneous group transformations (\ref{eq:tildeTrans}) with $\widetilde{m}_{ab} = \delta_{ab}$,
$\widetilde{\ell}_a \ne 0$, we have
\begin{align}
&{{[}0 a_{12} \ldots a_{1k}{]} {[}0 a_{22} \ldots a_{2k}{]} \ldots
[0 a_{t2} \ldots a_{tk}]}
{{[} b_{11} b_{12} \ldots b_{1,k\!-\!1}{]} \ldots {[} b_{r-t,1} b_{r2} \ldots b_{r-t,k\!-\!1}{]}0_1 0_2 \ldots 0_s} \nonumber\\
&\hspace{8em}\longrightarrow\nonumber\\
&{{[}0 a_{12} \ldots a_{1k}{]} {[}0 a_{22} \ldots a_{2k}{]} \ldots
[0 a_{t2} \ldots a_{tk}]}
{{[} b_{11} b_{12} \ldots b_{1,k\!-\!1}{]} \ldots {[} b_{r-t,1} b_{r2} \ldots b_{r-t,k\!-\!1}{]}0_1 0_2 \ldots 0_s} + \nonumber\\
&\widetilde{\ell}_{a_{12}} {{[}0 0 \ldots a_{1k}{]} {[}0 a_{22} \ldots a_{2k}{]} \ldots [0 a_{t2} \ldots a_{tk}]}
{{[} b_{11} b_{12} \ldots b_{1,k\!-\!1}{]} \ldots {[} b_{r-t,1} b_{r2} \ldots b_{r-t,k\!-\!1}{]}0_1 0_2 \ldots 0_s}+ \ldots  \nonumber\\
& + \widetilde{\ell}_{b_{11}}{{[}0 a_{12} \ldots a_{1k}{]} {[}0 a_{22} \ldots a_{2k}{]} \ldots [0 a_{t2} \ldots a_{tk}]} 
{{[} 0 b_{12} \ldots b_{1k\!-\!1}{]} \ldots {[} b_{r-t,1} b_{r2} \ldots b_{r-t,k\!-\!1}{]}0_1 0_2 \ldots 0_s}\nonumber\\
& +  \widetilde{\ell}_{b_{12}} {{[}0 a_{12} \ldots a_{1k}{]} {[}0 a_{22} \ldots a_{2k}{]} \ldots
[0 a_{t2} \ldots a_{tk}]} {{[} b_{11} 0 \ldots b_{1,k\!-\!1}{]} \ldots {[} b_{r-t,1} b_{r2} \ldots b_{r-t,k\!-\!1}{]}0_1 0_2 \ldots 0_s}\nonumber\\
&\hspace{20em}+ \ldots , \nonumber
\end{align}
wherein the coefficients of the $\widetilde{\ell}_{a_{\ldots}}$ terms vanish by anti-symmetry, but those of the
$\widetilde{\ell}_{b_{\ldots}}$ terms clearly do not. The components corresponding to the desired
$\overline{\lambda} \!=\! (r^{k\!-\!1})$ one-dimensional representation of $GL(k\!-\!1)$ within
$\lambda =  (r + s, r^{k-2}, t)$ is therefore \emph{not} invariant under inhomogeneous transformations corresponding to translations in $GL(k\!-\!1) \ltimes T(k\!-\!1) \cong GL_1(k)$ \emph{unless} the ${{[} b_{1,1}  \ldots b_{r-t,k\!-\!1}{]}}$ columns are absent, that is, $t \!\equiv\! r$.
Thus, the requirement of invariance of the one-dimensional representations under $GL_1(k)$ necessitates $\lambda = (r+s, r^{k\!-\!1})$ as claimed in part (ii) of Theorem~1.

\section{The construction of Markov invariants}\label{sec:YoungOps}
\setcounter{equation}{0}
\renewcommand{\theequation}{{B}-\arabic{equation}}

The standard construction of the irreducible modules $V^\lambda$ is given, for example, in \cite[Lecture 4]{fulton1991}. 
Here we modify this procedure, to give the explicit polynomial form of the Markov invariants.

Consider the representation of $\mathfrak{S}_m$ on $\otimes^m V$ defined by the action $v_1\otimes v_2\otimes \ldots v_m\mapsto v_{\alpha(1)}\otimes v_{\alpha(2)}\otimes \ldots \otimes v_{\alpha(m)}$ for all $\alpha\!\in\!\mathfrak{S}_m$. 
Given a \emph{standard} tableau $\mathfrak{T}$ with shape $\lambda$ and $|\lambda| = m$, define the permutations $p\in \mathfrak{S}_m$ as those that interchange the integers in the same row, and the permutations $q\in \mathfrak{S}_m$ as those that interchange numbers in the same column. 
In the algebra of the representation of the symmetric group whose action is defined above, consider the quantities
\beqn
A =\sum_{p\in \mathfrak{T}} p,\nonumber
\eqn
and
\beqn
B=\sum_{q\in\mathfrak{T}} \mbox{sign}(q)q.\nonumber
\eqn
The Young's operator corresponding to $\mathfrak{T}$ is then defined as
\beqn
Y^\lambda = BA.\nonumber
\eqn
It follows that for a standard tableau of shape $\lambda$, the corresponding Young's operator projects onto an irreducible module of $GL(k)$:
\beqn
V^\lambda=Y^\lambda\cdot \otimes^mV.\nonumber
\eqn
This construction is independent of $k$, and Young's operators corresponding to standard tableau of the same shape project onto equivalent modules. The independent tensor components of these irreducible modules are found by inserting integers from \emph{semi-standard} tableaux into the indices of the generic tensor. To compute the  explicit form of Markov invariants, we must apply this standard procedure to our special case.

Begin with the generic form of a monomial in the components of the tensor $\psi\!\in\!\otimes^m V$:
\beqn
\psi_{i_1\ldots i_m}\psi_{i_{m+1}\ldots i_{2m}}\ldots \psi_{i_{m(d-1)+1}\ldots i_{md}}.\nonumber
\eqn
To find the polynomial form of an invariant that arises from an inner product of Schur functions $\{\sigma_1\}\odot\{\sigma_2\}\odot\ldots\odot\{\sigma_m\}$ with $\sigma_a=\{r+s,r^{k\!-\!1}\}$ for all $1\leq a\leq m$, and $rk+s=d$, we must apply the Young's operators to these indices. In an abuse of notation we write
\beqn
\Psi_{i_1\ldots i_{dm}}:=Y^{\sigma_1}Y^{\sigma_2}\ldots Y^{\sigma_m}\psi_{i_1\ldots i_m}\psi_{i_{m+1}\ldots i_{2m}}\ldots \psi_{i_{m(d-1)+1}\ldots i_{md}},\nonumber
\eqn 
where each Young's operator $Y^{\sigma_a}$, $1\leq a\leq m$, is generated from a standard tableau of shape $\{\sigma_a\}$ with integers chosen from the set $\{a,m+a,\ldots,(d-1)m+a\}$. 
That is, each $Y^{\sigma_a}$ permutes the indices $i_a,i_{a+m},\ldots ,i_{a+md}$. The final step is to insert indices into $\Psi$ using the semi-standard tableau which results from filling the $1^{\text{st}}$ row with the integer 0, and, for $2\leq i\leq k$, the $i^\text{th}$ row with the integer $i$. 
The justification for filling the ``overhang'' of length $s$ in the first row of the tableau with the integer 0, is that in the basis given in Appendix~\ref{sec:Proofs}, the $\widetilde{\delta_0}$ component is an invariant subspace. 
For more details, including multiple examples, see \cite{sumner2006a}.

This procedure has been implemented to garner the polynomial form of the Markov invariants for phylogenetic trees with up to four leaves. 
These are presented in Table~\ref{tab:nomenclature}. 
The algorithms required were performed in \texttt{Mathematica} \cite{mathematica}, and, unfortunately, do not scale well for trees with more leaves. 
We are currently investigating the design of efficient algorithms for this construction, and note here that \cite{molev2007} provides a promising direction.

Additionally, in this construction the resulting polynomial form of the invariant is not in the $\delta_1,\delta_2,\ldots,\delta_k$ basis, and the required change of basis computation has thus far not been feasible. 
To evaluate the invariants on observed data, we therefore proceed by transforming the data itself into the appropriate basis. 
This allows us to evaluate Markov invariants on observed character pattern counts taken from phylogenetic data sets.

The calculation of unbiased forms, as defined in \S\ref{subsec:randomvariables}, is straight forward in principle. However, the calculation requires that the invariants be expressed in the $\delta_1,\delta_2,\ldots,\delta_k$ basis. This appears to be a rather challenging computational task, and to date the required algorithms have not been developed.

\section{Kimura 3ST model and phylogenetic invariants}
\label{sec:K3ST}

Our approach to phylogenetic models via group actions and representations finds specific application in some special cases, such as the Kimura 3ST \cite{kimura1981} model and certain generalizations to be described below. Here we provide a brief discussion as an illustration of our focus.

In the usual basis of point measures $\delta_A, \delta_C, \delta_U, \delta_G$, the K3ST rate matrix $Q$,
\begin{align}
	\left[\begin{array}{cccc}
	Q_{AA} & Q_{AG} & Q_{AU} & Q_{AC} \\
	Q_{GA} & Q_{GG} & Q_{GU} & Q_{GC} \\
	Q_{UA} & Q_{UG} & Q_{UU} & Q_{UC} \\
	Q_{CA} & Q_{CG} & Q_{CU} & Q_{CC} \end{array}\right] &= -(\alpha \!+\!\beta \!+\!
	\gamma) 1 +
	\left[ \begin{array}{cccc}
	0 & \alpha & \beta & \gamma \\
	\alpha & 0 & \gamma & \beta \\
	\beta & \gamma & 0 & \alpha \\
	\gamma & \beta & \alpha & 0 \end{array}\right]
\end{align}
can be re-written \cite{bashford2004},
\begin{align}
	Q=  & \, (\alpha \!+\!\beta \!+\!
	\gamma) \left(-1 + \frac{\alpha}{\alpha + \beta + \gamma} K_{\alpha}
	+ \frac{\beta}{\alpha + \beta + \gamma} K_{\beta} +
	\frac{\gamma}{\alpha + \beta + \gamma} K_{\gamma}\right),
\end{align}
where the three `Kimura matrices'
\begin{align}
	K_{\alpha} = \left[ \begin{array}{cccc}
	0 & 1 & 0 & 0 \\
	1 & 0 & 0 & 0\\
	0 & 0 & 0 & 1 \\
	0 & 0 & 1 & 0 \end{array}\right] , \quad
	K_{\beta} = \left[ \begin{array}{cccc}
	0 & 0 & 1 & 0 \\
	0 & 0 & 0 & 1\\
	1 & 0 & 0 & 0 \\
	0 & 1 & 0 & 0 \end{array}\right] , \quad
	K_{\gamma} = \left[ \begin{array}{cccc}
	0 & 0 & 0 & 1 \\
	0 & 0 & 1 & 0\\
	0 & 1 & 0 & 0 \\
	1 & 0 & 0 & 0 \end{array}\right] ,
\end{align}
span a Cartan (maximal commuting) sub\emph{algebra} of the group $SL(4)$, and therefore can be diagonalised simultaneously, via the well-known Hadamard transform \cite{hendy1993},
\begin{align}
	\label{eq:4by4similarity}
	& \begin{array}{rr}
	\quad{\mathsf H}=h\!\otimes\!h\!=\!\left[ \begin{array}{cccc}
	 1 &  1 &  1 &  1 \\
	 1 & -1 &  1 & -1 \\
	 1 &  1 & -1 & -1 \\
	 1 & -1 & -1 &  1 \end{array}\right] &, \quad
	{\mathsf H} K_{\alpha} {\mathsf H}^{-1} \!=\!\left[ \begin{array}{cccc}
	 1 & 0 & 0 & 0 \\
	0 & -1  & 0 & 0\\
	0 & 0 & 1 & 0 \\
	0 & 0 & 0 & -1 \end{array}\right], \nonumber \\ & \\
	{\mathsf H} K_{\beta} {\mathsf H}^{-1} \!=\! \left[ \begin{array}{cccc}
	1 & 0 & 0 & 0 \\
	0 & 1  & 0 & 0\\
	0 & 0 & -1 & 0 \\
	0 & 0 & 0 & -1 \end{array}\right] &, \quad
	{\mathsf H} K_{\gamma} {\mathsf H}^{-1} \!=\! \left[ \begin{array}{cccc}
	1 & 0 & 0 & 0 \\
	0 & -1  & 0 & 0\\
	0 & 0 & -1 & 0 \\
	0 & 0 & 0 & 1 \end{array}\right].\end{array}, \nonumber \\
&\mbox{with}
		\qquad {\mathsf h} = \left[ \begin{array}{cc} 1 & 1 \\
		1 & - 1 \end{array} \right].&
\end{align}
This simple observation means that under this model, rank-$m$ phylogenetic tensors have a spectral resolution given directly in terms of weights of the appropriate
${\times^m}(gl(1)\times gl(1)\times gl(1))$ abelian subalgebra of $\times^m GL(4)$ (equivalently the weight decomposition of the corresponding representation of $\times^m SL(4)$).

In fact, a stronger statement is possible. The action of group elements of the form $M(t)= e^{tQ}$ turns out to be covariant with respect to the operator $\delta$ introduced in \S\ref{sec:phylotens} above, describing branching in the general phylogenetic model -- explicitly, in the notation of \S\ref{sec:phylotens}, we have
\begin{align}
\label{eq:PullBack}
\delta \cdot \exp( a K_\alpha + b K_\beta + c K_\gamma) 
= & \, \exp( a K_\alpha \otimes K_\alpha  +
b K_\beta \otimes K_\beta+ c K_\gamma \otimes K_\gamma) \cdot \delta.
\end{align}

Applied to  a phylogenetic tensor $P$ with underlying arbitrary tree ${\mathcal T}$, (\ref{eq:PullBack}) then means that, under this model, the action of the Markov operators on each internal edge can be pulled back to the pendant edges, at the expense of a more complicated edge-mixing transformation. 
In final form, $P$ is given by the action of a certain element of
$(GL(1)\times GL(1)\times GL(1))^{\times m}$ within $GL(4^m)$, with the embedding fixed by the tree, applied to the maximally branched product measure $\delta^{(m-1)} \cdot\pi$, defined by
\[
\delta^{(m-1)} \cdot\pi = \sum_i \pi_i \delta_i \otimes \cdots \otimes \delta_i ,
\]
with $m$ tensor products in each term\footnote{This construction can be achieved by noting, for any linear operators $A,B,C,D$ with $AC\!=\!CA$ and $BD\!=\!DB$, algebraic identities like $e^A\otimes e^B=e^{(A\otimes 1+1\otimes B)}$, and $e^A\otimes e^B\cdot e^{C\otimes D}=e^{(A\otimes 1+1\otimes B+C\otimes D)}$.}.
Further details can be found in \cite{bashford2004}.

Analyses of this sort are useful both analytically, and in explicit calculations. In particular, the identification of phylogenetic invariants for given trees becomes straightforward, once the components of $P$ are written in the diagonal Hadamard basis. The group representation analysis provides a useful alternative to discrete Fourier transform methods which have been successfully applied where rate matrices admit a symmetry with respect to a discrete colour group, $Z_2 \times Z_2 \times \cdots$ \cite{hendy1993,hendy1994}, and may also be useful in the characterisation of phylogenetic varieties in the phylogenetic invariants analysis \cite{allman2004,strumfels2004} (see also the discussion in \S\ref{sec:Outlook}).

The above considerations generalize to the case of any $k$-state model wherein the off-diagonal part of the rate operator is a linear combination of a maximal set of commuting permutation matrices belonging to ${\mathfrak S}_{k}$, which guarantees (\ref{eq:PullBack}). For example, this class would include a 3-state model even simpler than the K3ST model, but which is non-symmetric, and for which the Hadamard basis is complex:
\begin{align}
Q=  & \,  (\alpha \!+\!\beta) \big(-1 + \frac{\alpha}{\alpha + \beta} K_{\alpha}
	+ \frac{\beta}{\alpha + \beta} K_{\beta} \big), \nonumber
\\	
	K_{\alpha} = & \, \left[ \begin{array}{ccc}
	0 & 0 & 1  \\
	1 & 0 & 0  \\
	0 & 1 & 0  \end{array}\right] , \qquad
	K_{\beta} =  \left[ \begin{array}{ccc}
	0 & 1 & 0  \\
	0 & 0 & 1  \\
	1 & 0 & 0  \end{array}\right].
\end{align}
See \cite{bashford2004} for further details

\end{appendix}

%% file: markovpleth.bbl
\begin{thebibliography}{10}

\bibitem{allman2003}
E.~S. Allman and J.~A. Rhodes.
\newblock Phylogenetic invariants of the general {Markov} model of sequence
  mutation.
\newblock {\em Math. Biosci.}, 186:113--144, 2003.

\bibitem{allman2004}
E.~S. Allman and J.~A. Rhodes.
\newblock Phylogenetic ideals and varieties for the general {M}arkov model.
\newblock {\em Adv. Appl. Math.}, to appear, 2007.

\bibitem{baker2003}
A.~Baker.
\newblock {\em Matrix Groups: An Introduction to Lie Group Theory}.
\newblock Springer-Verlag, 2003.

\bibitem{barry1987}
D.~Barry and J.~A. Hartigan.
\newblock Asynchronous distance between homologous {DNA} sequences.
\newblock {\em Biometrics}, 43:261--276, 1987.

\bibitem{bashford2004}
J.~D. Bashford, P.~D. Jarvis, J.~G. Sumner, and M.~A. Steel.
\newblock {$U(1)\times U(1)\times U(1)$} symmetry of the {K}imura 3{S}{T} model
  and phylogenetic branching processes.
\newblock {\em J. Phys. A Math. Gen.}, 37:L1--L9, 2004.

\bibitem{biedenharn1981}
L.~C. Biedenharn and J.~D. Louck.
\newblock {\em The Racah-Wigner Algebra in Quantum Theory}.
\newblock Addison-Wesley, 1981.

\bibitem{biedenharn1990}
L.~C. Biedenharn and J.~D. Louck.
\newblock Inhomogeneous basis set of symmetric polynomials defined by tableaux.
\newblock {\em Proc. Natl. Acad. Sci. U.S.A.}, 87:1441--1445, 1990.

\bibitem{bininda2004}
O.~R.~P. Bininda-Emonds, editor.
\newblock {\em Phylogenetic Supertrees: Combining Information to Reveal the
  Tree of Life}.
\newblock Springer, 2004.

\bibitem{bryant2005}
D.~Bryant.
\newblock On the uniqueness of the selection criterion in {N}eighbor-{J}oining.
\newblock {\em J. Class.}, 22:3--15, 2005.

\bibitem{bryant2005b}
D.~Bryant, N.~Galtier, and M.-A. Poursat.
\newblock Likelihood calculation in molecular phylogenetics.
\newblock In Olivier Gascuel, editor, {\em Mathematics of Evolution and
  Phylogenetics}, pages 33--62. Oxford University Press, 2005.

\bibitem{burnham2002}
K.~P. Burnham and D.~Anderson.
\newblock {\em Model Selection and Multi-Model Inference}.
\newblock Springer-Verlag, 2002.

\bibitem{casanellas2007}
M.~Casanellas and J.~Fern\'andez-S\'anchez.
\newblock Performance of a new invariants method on homogeneous and
  nonhomogeneous quartet trees.
\newblock {\em Mol. Biol. Evol.}, 24:288--293, 2007.

\bibitem{cavender1987}
J.~A. Cavender and J.~Felsenstein.
\newblock Invariants of phylogenies in a simple case with discrete states.
\newblock {\em J. Class.}, 4:57--71, 1987.

\bibitem{charleston2001}
M.~A. Charleston.
\newblock Hitch-hiking: A parallel heuristic search strategy, applied to the
  phylogeny problem.
\newblock {\em J. Comput. Biol.}, 8:79--91, 2001.

\bibitem{coffman2000}
V.~Coffman, J.~Kundu, and W.~K. Wootters.
\newblock Distributed entanglement.
\newblock {\em Phys. Rev. A}, 61(5):052306, Apr 2000.

\bibitem{dur2000}
W.~Dur, G.~Vidal, and J.~I. Cirac.
\newblock Three qubits can be entangled in two inequivalent ways.
\newblock {\em Phys. Rev. A}, 62:062314, 2000.

\bibitem{eriksson2008}
N.~Eriksson.
\newblock Using invariants for phylogenetic tree construction.
\newblock {\em \texttt{eprint arXiv:0709.2890}}, to appear.

\bibitem{eriksson2008b}
N.~Eriksson and Y.~Yao.
\newblock Metric learning for phylogenetic invariants.
\newblock {\em \texttt{eprint arXiv:q-bio/0703034}}, 2008.

\bibitem{evans1993}
S.~N. Evans and T.~P. Speed.
\newblock Invariants of some probability models used in phylogenetic inference.
\newblock {\em Annals of Statististics}, 21(1):355--377, 1993.

\bibitem{fauser2006}
B.~Fauser, P.~D. Jarvis, R.~C. King, and B.~G. Wybourne.
\newblock New branching rules induced by plethysm.
\newblock {\em J. Phys. A Math. Gen.}, 39:2611--2655, 2006.

\bibitem{felsenstein1978}
J.~Felsenstein.
\newblock Cases in which parsimony or compatibility methods will be positively
  misleading.
\newblock {\em Syst. Zool.}, 27:401--410, 1978.

\bibitem{felsenstein2004}
J.~Felsenstein.
\newblock {\em Inferring Phylogenies}.
\newblock Sinauer Associates, 2004.

\bibitem{phylip}
J.~Felsenstein.
\newblock {\em PHYLIP (Phylogeny Inference Package) version 3.6}.
\newblock Distributed by the author. Department of Genome Sciences, University
  of Washington, Seattle, 2005.

\bibitem{fulton1991}
W.~Fulton and J.~Harris.
\newblock {\em Representation Theory}.
\newblock Graduate Text in Mathematics. Springer-Verlag, 1991.

\bibitem{gascuel2005}
O.~Gascuel, editor.
\newblock {\em Mathematics of Evolution and Phylogenetics}.
\newblock Oxford University Press, 2005.

\bibitem{steel2006}
O.~Gascuel and M.~Steel.
\newblock Neighbor-{J}oining revealed.
\newblock {\em Mol. Biol. Evol.}, 23:1997--2000, 2006.

\bibitem{goodman1970}
G.~S. Goodman.
\newblock An intrinsic time for non-stationary finite {Markov} chains.
\newblock {\em Probab. Theor. Relat. Field.}, 16:165--180, 1970.

\bibitem{goodman1998}
R.~Goodman and N.~R. Wallach.
\newblock {\em Representations and Invariants of the Classical Groups}.
\newblock Cambridge University Press, 1998.

\bibitem{gu1996}
X.~Gu and W.~H. Li.
\newblock Bias-corrected paralinear and logdet distances and tests of molecular
  clocks and phylogenies under non-stationary nucleotide frequencies.
\newblock {\em Mol. Biol. Evol.}, 13:1375--1383, 1996.

\bibitem{halmos1974}
P.~R. Halmos.
\newblock {\em Measure Theory}.
\newblock Springer-Verlag, 1974.

\bibitem{hendy1993}
M.~D. Hendy and D.~Penny.
\newblock Spectral analysis of phylogenetic data.
\newblock {\em J. Class.}, 10:1--20, 1993.

\bibitem{hendy1994}
M.~D. Hendy, D.~Penny, and M.~Steel.
\newblock A discrete {F}ourier analysis for evolutionary trees.
\newblock {\em Proc. Natl. Acad. Sci.}, 91:3339--3343, 1994.

\bibitem{hillis1994}
D.~Hillis, J.~Huelsenbeck, and D.~Swofford.
\newblock Hobgoblin of phylogenetics?
\newblock {\em Nature}, 369:363--364, 1994.

\bibitem{hordijk2005}
W.~Hordijk and O.~Gascuel.
\newblock Improving the efficiency of {SPR} moves in phylogenetic tree search
  methods based on maximum likelihood.
\newblock {\em Bioinformatics}, 21:4338--4347, 2005.

\bibitem{huelsenbeck1995}
J.~P. Huelsenbeck.
\newblock Performance of phylogenetic methods in simulation.
\newblock {\em Syst. Biol.}, 44:17--48, 1995.

\bibitem{huelsenbeck2004}
J.~P. Huelsenbeck, B.~Larget, and M.~E. Alfaro.
\newblock Bayesian phylogenetic model selection using reversible jump {Markov}
  chain {Monte Carlo}.
\newblock {\em Mol. Biol. Evol.}, 21:1123--1133, 2004.

\bibitem{huson1999}
Daniel~H. Huson, Scott~M. Nettles, and Tandy~J. Warnow.
\newblock Disk-covering, a fast-converging method for phylogenetic tree
  reconstruction.
\newblock {\em J. Comput. Biol.}, 6:369--386, 1999.

\bibitem{isoifescu1980}
M.~Iosifescu.
\newblock {\em Finite Markov Processes and Their Applications}.
\newblock John Wiley and Sons, Chichester, 1980.

\bibitem{itzykson1980}
C.~Itzykson and J-B. Zuber.
\newblock {\em Quantum Field Theory}.
\newblock McGraw-Hill, New York, 1980.

\bibitem{jarvis2005}
P.~D. Jarvis, J.~D. Bashford, and J.~G. Sumner.
\newblock Path integral formulation and{ Feynman} rules for phylogenetic
  branching models.
\newblock {\em J. Phys. A Math. Gen.}, 38:9621--9647, 2005.

\bibitem{jayaswal2005}
V.~Jayaswal, L.~S. Jermiin, and J.~Robinson.
\newblock Estimation of phylogeny using a general {M}arkov model.
\newblock {\em Evolutionary Bioinformatics Online}, 1:62--80, 2005.

\bibitem{jayaswal2007}
V.~Jayaswal, J.~Robinson, and L.~Jermiin.
\newblock Estimation of phylogeny and invariant sites under the general
  {Markov} model of nucleotide sequence evolution.
\newblock {\em Syst. Biol.}, 56:155--162, 2007.

\bibitem{hetero}
L.~S. Jermiin, S.~Y.~W. Ho, F.~Ababneh, J.~Robinson, and A.~W.~D. Larkum.
\newblock Hetero: A program to simulate the evolution of {DNA} on four-taxon
  trees.
\newblock {\em Appl. Bioinformatics}, 2:159--163, 2003.

\bibitem{jermiin2004}
L.~S. Jermiin, S.~Y.~W. Ho, F.~Ababneh, J.~Robinson, and A.~W.~D. Larkum.
\newblock The biasing effect of compositional heterogeneity on phylogenetic
  estimates may be underestimated.
\newblock {\em Syst. Biol.}, 53:638--643, 2004.

\bibitem{jermiin2008}
L.~S. Jermiin, V.~Jayaswal, F.~Ababneh, and J.~Robinson.
\newblock Phylogenetic model evaluation.
\newblock In J.~Keith, editor, {\em Bioinformatics - Volume I: Data, Sequences
  Analysis and Evolution}, pages 331--363. Humana Press, Totowa, NJ, 2008.

\bibitem{johnson1985}
J.~E. Johnson.
\newblock {M}arkov-type {Lie} groups in {$GL(n,{R})$}.
\newblock {\em J. Math. Phys.}, 26:252--257, 1985.

\bibitem{kelarev2002}
A.~Kelarev.
\newblock {\em Ring Constructions and Applications}.
\newblock World Scientific, 2002.

\bibitem{keown1975}
R.~Keown.
\newblock {\em An Introduction to Group Representation Theory}.
\newblock Academic Press, New York, 1975.

\bibitem{kimura1981}
M.~Kimura.
\newblock Estimation of evolutionary distances between homologous nucleotide
  sequences.
\newblock {\em Proc. Natl. Acad. Sci.}, 78:1454–--1458, 1981.

\bibitem{king1975}
R.~C. King.
\newblock Branching rules for classical {Lie} groups using tensor and spinor
  methods.
\newblock {\em J. Phys. A Math. Gen.}, 8:429--449, 1975.

\bibitem{lake1987}
J.~A. Lake.
\newblock A rate-independent technique for analysis of nucleic acid sequences:
  evolutionary parsimony.
\newblock {\em Mol. Biol. Evol.}, 4:167--191, 1987.

\bibitem{lake1994}
J.~A. Lake.
\newblock Reconstructing evolutionary trees from {DNA} and protein sequences:
  Paralinear distances.
\newblock {\em Proceedings of the National Academy of Sciences}, 91:1455--1459,
  1994.

\bibitem{landsberg2006}
J.~M. Landsberg and L.~Manivel.
\newblock Generalizations of {S}trassen's equations for secant varieties of
  {S}egre varieties.
\newblock {\em Communications in Algebra}, 36:405--422, 2008.

\bibitem{littlewood1940}
D.~E. Littlewood.
\newblock {\em The Theory of Group Characters}.
\newblock Clarendon Press, Oxford, 1940.

\bibitem{littlewood1957}
D.~E. Littlewood.
\newblock Plethysm and the inner product of {S}-functions.
\newblock {\em J. Lond. Math. Soc.}, s1--32:18--22, 1955.

\bibitem{lockhart1996}
P.~J. Lockhart, A.~W.~D. Larkum, M.~A. Steel, P.~J. Waddell, and D.~Penny.
\newblock Evolution of chlorophyll and bacteriochlorophyll: The problem of
  invariant sites in sequence analysis.
\newblock {\em Proc. Natl. Acad. Sci. U.S.A.}, 93:1930--1943, 1996.

\bibitem{lockhart2006}
P.~J. Lockhart, P.~Novis, B.~G. Milligan, J.~Riden, A.~Rambaut, and A.~W.~D.
  Larkum.
\newblock Heterotachy and tree building: A case study with plastids and
  eubacteria.
\newblock {\em Mol. Biol. Evol.}, pages 40--45, 2006.

\bibitem{lockhart1998}
P.~J. Lockhart, M.~A. Steel, A.~C. Barbrook, D.~H. Huson, and C.~J. Howe.
\newblock A covariotide model describes the evolution of oxygenic
  photosynthesis.
\newblock {\em Mol. Biol. Evol.}, 15:1183--1188, 1998.

\bibitem{lockhart1994}
P.~J. Lockhart, M.~A. Steel, M.~D. Hendy, and D.~Penny.
\newblock Recovering evolutionary trees under a more realistic model of
  sequence evolution.
\newblock {\em Mol. Biol. Evol.}, 11:605--612, 1994.

\bibitem{macdonald1979}
I.~G. MacDonald.
\newblock {\em Symmetric Functions and Hall Polynomials}.
\newblock Clarendon Press, Oxford, 1979.

\bibitem{massingham2007}
T.~Massingham and N.~Goldman.
\newblock Statistics of the log-det estimator.
\newblock {\em MBE Advance Access published August 16, 2007}, 2007.

\bibitem{matsen2006}
F.~A. Matsen and S.~N. Evans.
\newblock Ubiquity of synonymity: Almost all large binary trees are not
  uniquely identified by their spectra or their immanantal polynomials.
\newblock {\em \texttt{eprint arXiv:q-bio/0512010}}, 2006.

\bibitem{mccullagh1987}
P.~McCullagh.
\newblock {\em Tensor Methods in Statistics}.
\newblock Chapman and Hall, 1987.

\bibitem{molev2007}
A.~Molev.
\newblock On the fusion procedure for the symmetric group.
\newblock {\em \texttt{eprint arXiv:math/0612207}}, 2007.

\bibitem{mourad2004}
B.~Mourad.
\newblock On a {Lie}-theoretic approach to generalised doubly stochastic
  matrices and applications.
\newblock {\em Linear and Multilinear algebra}, 52:99--113, 2004.

\bibitem{olver2003}
P.~J. Olver.
\newblock {\em Classical Invariant Theory}.
\newblock Cambridge University Press, Cambridge, 2003.

\bibitem{pagel2004}
M.~Pagel and A.~Meade.
\newblock A phylogenetic mixture model for detecting pattern-heterogeneity in
  gene sequence or character-state data.
\newblock {\em Syst. Biol.}, 53:571--581, 2004.

\bibitem{ape}
E.~Paradis, J.~Claude, and K.~Strimmer.
\newblock {APE}: analyses of phylogenetics and evolution in {R} language.
\newblock {\em Bioinformatics}, 20:289--290, 2004.

\bibitem{penny2001}
D.~Penny, B.~J. McComish, M.~A. Charleston, and M.~D. Hendy.
\newblock Mathematical elegance with biochemical realism: the covarion model of
  molecular evolution.
\newblock {\em J. Mol. Evol.}, 53:711--723, 2001.

\bibitem{posada2004}
D.~Posada and T.~R. Buckley.
\newblock Model selection and model averaging in phylogenetics: advantages of
  {Akaike} information criterion and {Bayesian} approaches over likelihood
  ratio tests.
\newblock {\em Syst. Biol.}, 53:793--–808, 2004.

\bibitem{Rproject}
{R Development Core Team}.
\newblock {\em R: A Language and Environment for Statistical Computing}.
\newblock R Foundation for Statistical Computing, Vienna, Austria, 2006.

\bibitem{semple2003}
C.~Semple and M.~Steel.
\newblock {\em Phylogenetics}.
\newblock Oxford Press, 2003.

\bibitem{seshadri1962}
C.~S. Seshadri.
\newblock On a theorem of {Weitzenb\"{o}ck} in invariant theory.
\newblock {\em J. Math. Kyoto. Univ.}, 1:403--409, 1962.

\bibitem{steel2002}
M.~Steel.
\newblock Some statistical aspects of the maximum parsimony method.
\newblock In R.~DeSalle, G.~Giribet, and W.~Wheeler, editors, {\em Molecular
  Systematics and Evolution: Theory and Practice}, pages 125--140.
  {Birkh\"{a}user} Verlag, 2002.

\bibitem{steel2005}
M.~Steel.
\newblock Should phylogenetic models be trying to 'fit an elephant'?
\newblock {\em Genetics}, 21:307--309, 2005.

\bibitem{steel1994}
M.~A. Steel.
\newblock Recovering a tree from the leaf colourations it generates under a
  {Markov} model.
\newblock {\em Appl. Math. Lett.}, 7:19--24, 1994.

\bibitem{steel1993}
M.~A. Steel, L.~Szekely, P.~L. Erdos, and P.~Waddell.
\newblock A complete family of phylogenetic invariants for any number of taxa
  under {K}imura's {3ST} model.
\newblock {\em N.Z. J. Bot.}, 31:289--296, 1993.

\bibitem{strimmer1996}
K.~Strimmer and A.~von Haeseler.
\newblock Quartet puzzling: A quartet maximum likelihood method for
  reconstructing tree topologies.
\newblock {\em Mol. Biol. Evol.}, 13:964--960, 1996.

\bibitem{sturmfels2007}
B.~Sturmfels.
\newblock Open problems in algebraic statistics.
\newblock In {\em M. Putinar and S. Sullivant ({E}ds.), Emerging Applications
  of Algebraic Geometry, I.M.A. Volumes in Mathematics and its Applications},
  to appear.

\bibitem{strumfels2004}
B.~Sturmfels and S.~Sullivant.
\newblock Toric ideals of phylogenetic invariants.
\newblock {\em J. Comput. Biol.}, 12:204--228, 2005.

\bibitem{sumner2006a}
J.~G. Sumner.
\newblock {Entanglement, Invariants, and Phylogenetics}.
\newblock {\em PhD thesis, University of Tasmania,
  \texttt{http://eprints.utas.edu.au}}, 2006.

\bibitem{squangle}
J.~G. Sumner.
\newblock Phylogenetic quartet inference using the squangles.
\newblock {\em University of Sydney,
  \texttt{http://www.it.usyd.edu.au/\~{}mcharles/software}}, 2008.

\bibitem{sumner2005}
J.~G. Sumner and P.~D. Jarvis.
\newblock Entanglement invariants and phylogenetic branching.
\newblock {\em J. Math. Biol.}, 51:18--36, 2005.

\bibitem{sumner2006}
J.~G. Sumner and P.~D. Jarvis.
\newblock Using the tangle: A consistent construction of phylogenetic distance
  matrices.
\newblock {\em Math. Biosci.}, 204:49--67, 2006.

\bibitem{tuffley1997}
C.~Tuffley and M.~A. Steel.
\newblock Links between maximum likelihood and maximum parsimony under a simple
  model of site substitution.
\newblock {\em Bull. Math. Biol.}, 59:581--607, 1997.

\bibitem{weitzenbock1931}
R.~{Weitzenb\"{o}ck}.
\newblock {\"{U}ber} die {Invarianten} von linearen {Gruppen}.
\newblock {\em Acta. Math.}, 58:231--293, 1931.

\bibitem{weyl1950}
H.~Weyl.
\newblock {\em The Theory of Groups and Quantum Mechanics}.
\newblock Dover Publications, 1950.

\bibitem{whippman1965}
M.~L. Whippman.
\newblock Branching rules for simple {Lie} groups.
\newblock {\em J. Math. Phys.}, 6:1534--1539, 1965.

\bibitem{wilkinson2006}
M.~Wilkinson and J.~A. Cotton.
\newblock Supertree methods for building the tree of life: Divide-and-conquer
  approaches to large phylogenetic problems.
\newblock In T.~R. Hodkinson and J.~A.~N. Parnell, editors, {\em Reconstructing
  the Tree of Life: Taxonomy and Systematics of Species Rich Taxa. Systematics
  Association Special Volume 72}. CRC Press, 2006.

\bibitem{mathematica}
{Wolfram Research, Inc.}
\newblock \texttt{Mathematica 5.2}.
\newblock 2005.

\bibitem{schur}
B.~G. Wybourne.
\newblock \texttt{Schur}: An interactive programme for calculating properties
  of {Lie} groups. version 6.03.
\newblock {\em \texttt{http://sourceforge.net/projects/schur}}, 2004.

\bibitem{yang1994}
Z.~Yang.
\newblock Maximum likelihood phylogenetic estimation from {DNA} sequences with
  variable rates over sites: approximate methods.
\newblock {\em J. Mol. Evol.}, 39:306--314, 1994.

\bibitem{yang2006}
Z.~Yang.
\newblock {\em Computational Molecular Evolution}.
\newblock Oxford University Press, 2006.

\bibitem{zharkikh1994}
A.~Zharkikh.
\newblock Estimation of evolutionary distance between nucleotide sequences.
\newblock {\em J. Mol. Evol.}, 39:315--329, 1994.

\end{thebibliography}
